\documentclass[12pt,a4paper]{article}
\usepackage[utf8]{inputenc}
\usepackage{bm}
\usepackage{color}
\usepackage{amsmath}
\usepackage{amsthm}
\usepackage{amssymb}
\usepackage{graphicx}
\usepackage{verbatim}
\usepackage[breaklinks]{hyperref}

\newtheorem{lemma}{Lemma}
\newtheorem{theorem}{Theorem}

\newtheorem{claim}{Claim}

\newtheorem*{mainlemma}{Main Lemma}

\theoremstyle{definition}
\newtheorem{definition}{Definition}

\newcommand{\set}[2]{\{ \, #1 \mid #2 \, \}}

\renewcommand{\epsilon}{\varepsilon}

\begin{document}
\sloppy

\title{Nondeterministic tree-walking automata are not closed under complementation}
\author{Olga Martynova\thanks{%
	Department of Mathematics and Computer Science,
	St.~Petersburg State University, 7/9 Universitetskaya nab., Saint Petersburg 199034, Russia.
	E-mail: \texttt{olga22mart@gmail.com}
} \and Alexander Okhotin\thanks{%
	Department of Mathematics and Computer Science,
	St.~Petersburg State University, 7/9 Universitetskaya nab., Saint Petersburg 199034, Russia.
	E-mail: \texttt{alexander.okhotin@spbu.ru}
}}

\maketitle

\begin{abstract}
It is proved that the family of tree languages
recognized by nondeterministic tree-walking automata
is not closed under complementation,
solving a problem raised by Boja\'nczyk and Colcombet
(\href{https://doi.org/10.1137/050645427}{``Tree-walking automata do not recognize all regular languages''},
\emph{SIAM J.\ Comp.} 38 (2008) 658--701).
In addition, it is shown that nondeterministic tree-walking automata
are stronger than unambiguous tree-walking automata.
\end{abstract}

\section{Introduction}

Tree-walking automata, first studied by Aho and Ullman~\cite{AhoUllman},
are among the fundamental models in automata theory.
A tree-walking automaton walks over a labelled input tree of bounded degree, following the edges;
at each moment, an automaton is at some node and is in one of finitely many states,
and it uses its transition function to decide which edge to follow and which state to enter.
The main questions about the expressive power of tree-walking automata
were open for several decades,
until the breakthrough results of Boja\'nczyk and Colcombet~\cite{BojanczykColcombet_det,BojanczykColcombet_reg},
who proved that deterministic tree-walking automata (DTWA) are
weaker than nondeterministic tree-walking automata (NTWA),
which are in turn weaker than bottom-up tree automata.

In these papers, Boja\'nczyk and Colcombet
presented a list of three main problems on tree-walking automata:
two of them are the problems they solved,
and the third problem is whether the class of tree languages recognized by NTWA
is closed under complementation.
This paper gives a negative solution to this problem,
presenting a tree language recognized by an NTWA,
such that its complement cannot be recognized by any NTWA.

Research on tree-walking automata and related models has been conducted in several directions.
In particular, tree-walking automata with pebbles
were introduced by Engelfriet and Hoogeboom~\cite{EngelfrietHoogeboom},
who proved them to be at most as powerful as bottom-up tree automata.
Later, Boja\'nczyk et al.~\cite{BojanczykSamuelidesSchwentickSegoufin}
proved strict hierarchies in the number of pebbles
for both deterministic and nondeterministic pebble tree-walking automata.
They also proved that no number of pebbles can help deterministic automata
to simulate NTWA without pebbles.
Logical characterizations of pebble tree-walking automata
were given by Engelfriet and Hoogeboom~\cite{EngelfrietHoogeboom,EngelfrietHoogeboom_capture_logic}
and by Neven and Schwentick~\cite{NevenSchwentick}.

For deterministic tree-walking automata,
every automaton can be transformed to one that halts on every tree:
this was first done by Muscholl et al.~\cite{MuschollSamuelidesSegoufin}
using Sipser's~\cite{Sipser_halting} method
of traversing the tree of all computations ending in the accepting configuration.
This result also implies the closure of DTWA under complementation.
Later, Kunc and Okhotin~\cite{KuncOkhotin_reversible}
presented a generalized construction applicable to graph-walking automata
and producing reversible automata,
and also reduced the number of states in the resulting automata from quadratic to linear
in the size of the given deterministic automaton.

Much is known about the complexity of decision problems for tree-walking automata.
The emptiness and the inclusion problems for both DTWA and NTWA are EXPTIME-complete,
see Boja\'nczyk~\cite{Bojanczyk},
in contrast to the P-complete emptiness problem
for the more powerful bottom-up tree automata, see Veanes~\cite{Veanes}.
Samuelides and Segoufin~\cite{SamuelidesSegoufin}
determined the complexity of the emptiness and the inclusion problems
for $k$-pebble tree-walking automata: they are $k$-EXPTIME-complete.
Among the recent results,
Boja\'nczyk~\cite{Bojanczyk_separating_by_dtwa} proved
that it is undecidable whether two regular tree languages
can be separated by a deterministic tree-walking automaton.
For graph-walking automata, both deterministic and nondeterministic, Martynova~\cite{Martynova_emptiness}
proved that their non-emptiness problem is NEXPTIME-complete.

The second result of this paper is about \emph{unambiguous tree-walking automata (UTWA)}:
these are NTWA, which, for every tree they accept,
have a unique accepting computation, while the number of rejecting computations is unrestricted.
The unambiguous case has been studied for different models of automata
and for different complexity classes,
see the survey by Colcombet~\cite{Colcombet_unambiguity}.
Unambiguous automata are an intermediate model between deterministic and nondeterministic ones.
In particular, for finite automata on strings,
all three types of automata are equal in power,
and their relative succinctness has been a subject of much research:
see, e.g., the most recent contributions by Indzhev and Kiefer~\cite{IndzhevKiefer}
and G\"o\"os et al.~\cite{GoosKieferYuan}.
On the other hand, for tree-walking automata,
DTWA are weaker than NTWA~\cite{BojanczykColcombet_det},
and unambiguous tree-walking automata may theoretically coincide in power with either DTWA or NTWA,
or they may be strictly between them.
In this paper, we prove that UTWA are weaker than NTWA,
while the question of whether they are stronger than DTWA or not remains open.

\section{Trees and tree-walking automata}\label{section_definitions}

The notion of tree-walking automata is standard,
even though it can be presented in various notation.
This paper generally adopts the notation used by Boja\'nczyk and Colcombet~\cite{BojanczykColcombet_reg},
with insignificant modifications.

\begin{definition}
Let $\Sigma$ be an alphabet of labels.
A \emph{binary tree} over $\Sigma$ is a partial mapping $t \colon V \to \Sigma$,
where $V \subset \{1, 2\}^*$ is a finite non-empty and prefix-closed set of \emph{nodes},
and $t$ defines the \emph{label} of each node.
The empty string $\epsilon \in V$ is the \emph{root node},
and for each node $v \in V$, if the node $v1$ is in $V$,
it is called the \emph{left child} of $v$, and $v$ is its \emph{parent};
similarly, if $v2$ is in $V$,
it is the \emph{right child} of $v$.
A node either has both children or none;
in the latter case it
is called a \emph{leaf}.

The edge between a parent and a child
is defined by a function $d \colon V \times V \to D$,
where $D=\{+1, -1, +2, -2\}$ is the set of direction labels.
For every two nodes $u$ and $v$,
such that $v$ is the $i$-th child of $u$,
let $d(u, v)=+i$ and $d(v, u)=-i$.
The function $d$ is undefined on all other pairs.

A node $u \in V$ is said to be \emph{above} a node $v \in V$ if $v=uw$ for some $w \in \{1, 2\}^+$;
in this case, $v$ is said to be \emph{below} $u$.
A node $u$ is \emph{to the left} of $v$ if none of them is above the other,
and $u$ is lexicographically less than $v$;
in this case, $v$ is \emph{to the right} of $u$.

For each node $u \in V$, the \emph{subtree} of $t$ rooted at $u$
is a tree $t_u$ defined by $t_u(v)=t(uv)$ for all $v \in \{1, 2\}^*$ with $t(uv)$ defined.
\end{definition}

Some further notation for the trees turns out to be useful.
The following function describes the local position of a node in a tree.

\begin{definition}
Let $t$ be a tree with the set of nodes $V$.
Each node $v \in V$ is assigned a \emph{type},
drawn from the set
$\mathrm{Types}=\{\mathrm{root}, 1, 2\} \times \{\mathrm{internal}, \mathrm{leaf}\}$.
Define a function $\mathrm{Type} \colon V \to \mathrm{Types}$
by $\mathrm{Type}(v)=(x, y)$,
where $x$ defines whether $v$ is a left child, a right child or the root,
and $y$ specifies if $v$ is a leaf or not.
\begin{equation*}
	x=\begin{cases}
		\mathrm{root},
			& \text{if } v=\epsilon \\
		1,
			& \text{if } v=u1 \\
		2,
			& \text{if } v=u2
	\end{cases}
	\quad\quad
	y=\begin{cases}
		\mathrm{internal},
			& \text{if } v1, v2 \in V \\
		\mathrm{leaf},
			& \text{if } v1, v2 \notin V
	\end{cases}
\end{equation*}
\end{definition}

A (nondeterministic) tree-walking automaton is typically defined as follows.
It starts at the root in one of the initial states.
At each step it knows the current state and sees the label and the type of the current node.
Then, according to the transition function, it nondeterministically decides
to proceed to its parent or any of its children,
and changes its state.
If the automaton ever comes to the root in an accepting state, it accepts.

The definition assumed in this paper follows Boja\'nczyk and Colcombet~\cite{BojanczykColcombet_reg}.
Accordingly, the automaton is invested
with the knowledge of the label and the type of the destination node,
which are also part of a transition.
If the destination node does not have the specified type, that transition cannot be applied.
This ability does not give an automaton any extra power, since it could move to 
the destination node and backwards to collect the same information before making such a transition.
This way, transitions become symmetric and can be reversed.

\begin{definition}
A \emph{nondeterministic tree-walking automaton (NTWA)}
is a quintuple $A=(\Sigma, Q, Q_0, \delta, F)$,
where
\begin{itemize}
\item
$\Sigma$ is a finite alphabet of labels,
\item
$Q$ is a finite set of states,
\item
$Q_0 \subseteq Q$ is the set of initial states,
\item
$\delta \subseteq (Q \times \Sigma \times \mathrm{Types})^2 \times D$
is the transition relation,
\item
and $F \subseteq Q$ is the set of accepting states.
\end{itemize}

\emph{Configurations} of the automaton on a tree $t$ with a set of nodes $V$
are pairs $(q, v)$, with $q \in Q$ and $v \in V$.
A \emph{computation} from a configuration $(p, u)$ to a configuration $(q, v)$
is any sequence of the form $(r_0, w_0), (r_1, w_1), \ldots, (r_N, w_N)$,
where $N \geqslant 0$,
$r_i \in Q$ and $w_i \in V$ for all $i$,
$(r_0, w_0)=(p, u)$,
$(r_N, w_N)=(q, v)$,
and 
for every $i$, with $0 \leqslant i \leqslant N-1$,
the configurations $(r_i, w_i)$ and $(r_{i+1},w_{i+1})$ are connected by a transition,
that is,
\begin{equation*}
	\big(r_i, t(w_i), \mathrm{Type}(w_i),
	r_{i+1}, t(w_{i+1}), \mathrm{Type}(w_{i+1}),
	d(w_i, w_{i+1})\big) \in \delta.
\end{equation*}

An \emph{accepting computation} is a computation from an initial configuration $(q_0, v_0)$,
where $q_0 \in Q_0$ is an initial state and $v_0=\epsilon$ is the root node,
to any accepting configuration of the form $(q, v_0)$, with $q \in F$.
Thus, the automaton accepts only in the root.

A tree $t$ is accepted by the automaton $A$
if there is at least one accepting computation on $t$.
The language recognized by $A$, denoted by $L(A)$,
is the set of all trees the automaton accepts.
\end{definition}

\section{Separating language}\label{section_separating_language}

We consider binary trees,
with nodes labelled with $a$ or $b$, where $b$ is the blank symbol.
Let all leaves labelled with $a$ in a tree
be enumerated from left to right, starting with $1$,
and denoted by $u_1, u_2, \ldots$

\begin{figure}[t]
	\centerline{\includegraphics[scale=1]{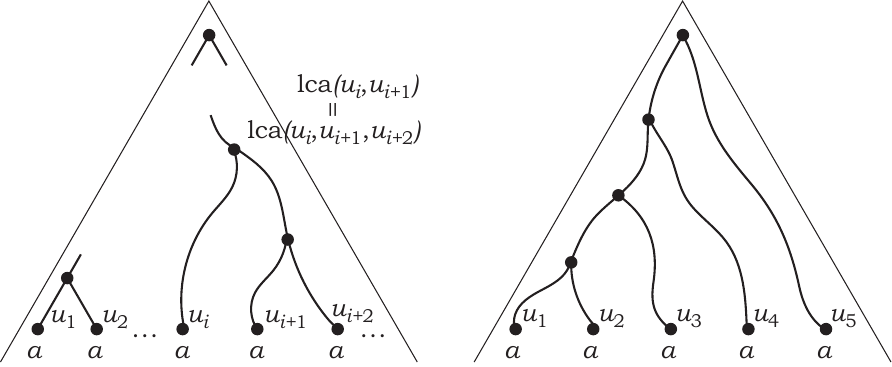}}
	\caption{(left) A tree in $L$; (right) a tree not in $L$.}
	\label{f:language_L}
\end{figure}

The language $L$ is defined as the set of all trees
in which there is a triple of leaves labelled with $a$,
with consecutive numbers $i$, $i+1$, $i+2$,
satisfying the condition
$\mathrm{lca}(u_i,u_{i+1},u_{i+2}) = \mathrm{lca}(u_i,u_{i+1})$,
where lca denotes the lowest common ancestor.
The form of trees satisfying this condition is illustrated in Figure~\ref{f:language_L}(left),
whereas trees that are not in $L$ have the form shown in Figure~\ref{f:language_L}(right).
In the figure, blank leaves are omitted along with paths leading to them.

The following theorem is the main result of this paper.

\begin{theorem}\label{ntwa_rec_L_but_not_compl_of_L}
There is a nondeterministic tree-walking automaton that recognizes the language $L$.
No nondeterministic tree-walking automaton can recognize the complement of $L$.
\end{theorem}

\begin{figure}[t]
	\centerline{\includegraphics[scale=1]{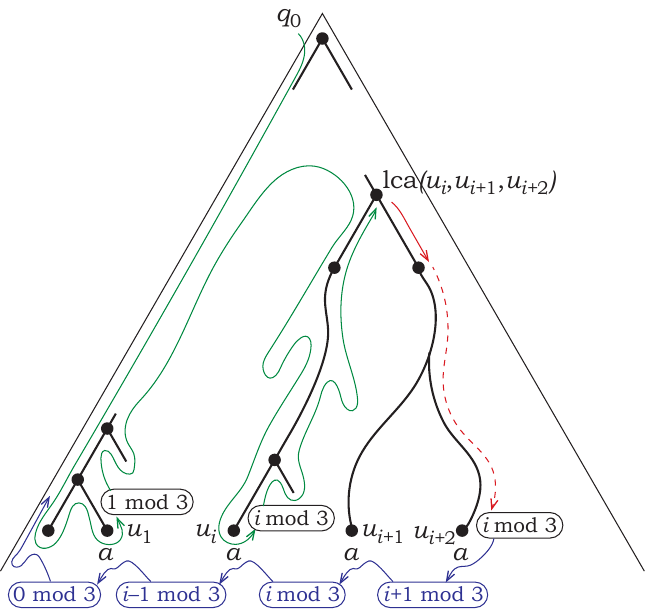}}
	\caption{NTWA accepting a tree from $L$, executing the algorithm from Section~\ref{section_separating_language}.}
	\label{f:L_recognizing_NTWA}
\end{figure}

In this section, the first part of the theorem is proved:
an NTWA $A_L$ recognizing the language $L$ will be described.
This NTWA works using the following algorithm,
illustrated in Figure~\ref{f:L_recognizing_NTWA}.
It begins with the traversal of the tree using depth-first search from left to right,
in which it counts modulo $3$ the number of leaves with label $a$.
At some point, while making a turn
(that is, while climbing from a left child to a parent node
and then immediately descending into its right child),
the automaton nondeterministically decides
that this node must be the lowest common ancestor
of a suitable triple of leaves $(u_i, u_{i+1}, u_{i+2})$,
and that the last $a$-labelled leaf encountered was $u_i$.
At this moment the automaton holds the residue $i \bmod 3$ in its state.
Next, the automaton, having descended into the right subtree,
nondeterministically guesses the path to the leaf $u_{i+2}$.
It ends its descent in some leaf labelled with $a$ (rejecting otherwise),
still remembering the residue $i \bmod 3$.
And now the automaton wants to check that the number of the current leaf is equivalent to $i+2$ modulo $3$.
To this end, the automaton starts another depth-first search, now from right to left,
initially keeping in memory the number $(i+1) \bmod 3$,
and decrementing it by one modulo $3$ at each leaf labelled with $a$.
If it finishes the traversal with residue zero in memory, then it accepts.

For every tree in the language $L$
there is an accepting computation of the automaton,
in which it just guesses the triple of leaves and their lowest common ancestor correctly,
and then guesses the path to $u_{i+2}$, as described above.

\begin{figure}[t]
	\centerline{\includegraphics[scale=1]{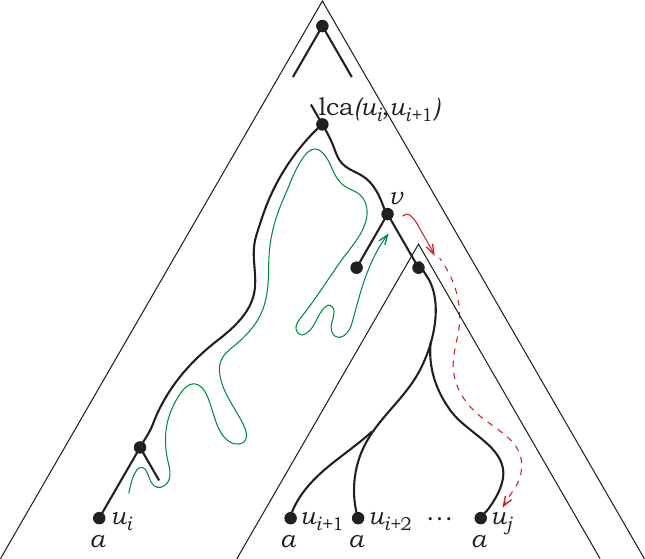}}
	\caption{The general form of accepting computation of NTWA $A_L$.}
	\label{f:L_recognizing_NTWA_2}
\end{figure}

Conversely, assume that the automaton accepts some tree.
It should be proved that this tree is in $L$.
At some moment, the automaton assumes that some node $v$
is the lowest common ancestor of three suitable leaves.
Let $u_i$ be the last leaf labelled with $a$ visited by the automaton up to this moment.
Then, either the leaf $u_i$ is in the left subtree of the node $v$,
or $u_i$ is in the left subtree of the lowest common ancestor of $u_i$ and $v$,
whereas $v$ is accordingly in the right subtree of this ancestor
(the latter case is illustrated in Figure~\ref{f:L_recognizing_NTWA_2}).
The automaton continues its computation by descending into the right subtree of $v$
and then nondeterministically chooses an $a$-labelled leaf in this subtree.
This leaf is $u_j$, with $j>i$,
because all leaves with numbers up to $i$ were traversed before visiting $v$.
Therefore, all leaves from $u_{i+1}$ to $u_j$ are in the right subtree of $v$.
Finally, the automaton checks that the number of the leaf it has found
has the same residue modulo $3$ as $i+2$.
This implies that $j \geqslant i+2$,
and that both leaves $u_{i+1}$ and $u_{i+2}$ lie in the right subtree of $v$.
Then the lowest common ancestor of $u_i$ and $u_{i+1}$
is precisely the lowest common ancestor of $u_i$ and $v$,
which coincides with the lowest common ancestor of the three nodes $u_i$, $u_{i+1}$ and $u_{i+2}$.

\section{Tools from the paper by Boja\'nczyk and Colcombet and their further properties}
\label{section_methods_BojanczykColcombet}

So the language $L$ is recognized by a nondeterministic tree-walking automaton.
Now it will be shown that no NTWA recognizes the complement of $L$.
Let $A$ be an NTWA, let $Q$ be its set of states.
The plan is to construct two trees, one not in $L$ and the other in $L$,
so that if the automaton $A$ accepts the former tree,
then it also accepts the latter tree.
Then the automaton $A$ cannot recognize the complement of the language $L$.

In constructing these trees and proving that the automaton operates on them as desired,
we use the tools developed by Boja\'nczyk and Colcombet~\cite{BojanczykColcombet_reg}.
Namely, the desired two trees are constructed out of the elements defined in their paper,
and we also use the basic properties of those elements in our proofs.
In this section, the required definitions and lemmata
by Boja\'nczyk and Colcombet~\cite{BojanczykColcombet_reg}
are presented,
along with several new lemmata addressing some further basic properties of those elements.

Following Boja\'nczyk and Colcombet~\cite{BojanczykColcombet_reg},
we consider trees with designated holes (ports)
and call them \emph{patterns}.

\begin{definition}
A \emph{pattern} is a binary tree with labels $\{a,b,*\}$,
in which the labels $a$ and $*$ may only be used in leaves,
and all leaves labelled with $*$ are left children.
A pattern must have at least two nodes.

The root of the tree is called the \emph{root port} or \emph{port $0$}.
All leaves labelled with $*$ are \emph{leaf ports},
enumerated from left to right starting with one.
The number of leaf ports in a pattern is called its \emph{rank}.
A pattern of rank $k$, with $k \geqslant 0$,
thus has the set of ports $\{0, 1, \ldots, k\}$,
and it is called a \emph{$k$-ary pattern}.
\end{definition}

Patterns can be attached to each other
by substituting one pattern for a leaf labelled with $*$ in another pattern.
This operation is called \emph{composition} of patterns,
and is denoted by $\Delta[\Delta_1, \ldots,\Delta_k]$:
here patterns $\Delta_1, \ldots,\Delta_k$ are attached to the leaf ports of a $k$-ary pattern $\Delta$.
If patterns are attached not to all leaf ports,
then $*$ is written instead of a pattern to be substituted.

Consider how an automaton can move through a pattern
if this pattern is a part of some tree.

\begin{definition}[Boja\'nczyk and Colcombet~{\cite[Defn.~3]{BojanczykColcombet_det}}, {\cite[Defn.~3]{BojanczykColcombet_reg}}]
Let $A$ be an NTWA with a set of states $Q$,
and let $\Delta$ be a pattern of rank $k$.
Let $p,q \in Q$ be two states and let $i,j \in \{0, 1, \ldots, k\}$ be two ports.
A computation of $A$ on $\Delta$
that begins in state $p$ in port $i$
and ends in state $q$ in port $j$,
without visiting any ports on the way,
and treating ports $i$ and $j$
as left non-leaf children labelled with $b$,
is said to be \emph{a run of type $(p,i,q,j)$}.

The automaton's \emph{transition relation} over $\Delta$
is $\delta_{\Delta} \subseteq Q \times \{0, 1, \ldots, k\} \times Q \times \{0, 1, \ldots, k\}$,
and it contains the types of all runs of $A$ over $\Delta$
(and no other quadruples).

Two patterns $\Delta$ and $\Delta'$ are called \emph{equivalent} (with respect to $A$)
if they are of the same rank
and their transition relations coincide: $\delta_{\Delta} = \delta_{\Delta'}$.
\end{definition}

A computation of zero length is considered as a run,
that is, $(p,i,p,i) \in \delta_{\Delta}$,
for all $p \in Q$ and $i \in \{0,1, \ldots, k\}$.

The equivalence relation on patterns is defined so that it respects composition:
if some pattern $\Delta$ is obtained as a composition of other patterns,
and if one of those subpatterns is replaced with an equivalent pattern,
then the resulting pattern will be equivalent to $\Delta$.

Most trees constructed in the papers
by Boja\'nczyk and Colcombet~\cite{BojanczykColcombet_det,BojanczykColcombet_reg}
are obtained by combining several specifically constructed patterns,
defined with respect to an automaton $A$,
which have the following remarkable properties.

\begin{lemma}[Boja\'nczyk and Colcombet~{\cite[Lemma 9]{BojanczykColcombet_det}}, {\cite[Lemma 3.1]{BojanczykColcombet_reg}}]
\label{lemma_delta_123}
Let $A$ be a tree-walking automaton.
Then there are patterns $\Delta_0$, $\Delta_1$, $\Delta_2$
of rank $0$, $1$ and $2$, respectively,
which have no labels $a$,
such that every pattern $\Delta$ of rank at most $2$
obtained as a composition of any number of patterns $\Delta_0$, $\Delta_1$, $\Delta_2$
is equivalent to one of $\Delta_0$, $\Delta_1$, $\Delta_2$
(the one of the same rank as $\Delta$).
\end{lemma}

In the following, $\Delta_0$, $\Delta_1$, $\Delta_2$
are patterns constructed for the automaton $A$ by Lemma~\ref{lemma_delta_123}.
One more basic pattern is $\Delta_a$, defined as $\Delta_1[\Delta'_a]$,
where $\Delta'_a$ is a tree with three nodes:
the root and the right leaf are labelled with $b$, 
whereas the left leaf has a label $a$,
as in the paper by Boja\'nczyk and Colcombet~\cite{BojanczykColcombet_reg}.

In this paper, the patterns $\Delta_0$, $\Delta_1$, $\Delta_2$ and $\Delta_a$,
as well as some combinations of a few such patterns,
shall be called \emph{elements},
as all trees considered in this paper shall be constructed out of them.

In patterns composed of elements $\Delta_0$, $\Delta_1$, $\Delta_2$,
one can attach an element $\Delta_1$ to any port and get an equivalent pattern.

\begin{lemma}[Boja\'nczyk and Colcombet~{\cite[Lemma 8]{BojanczykColcombet_det}}, {\cite[Fact 3.2]{BojanczykColcombet_reg}}]
\label{lemma_attaching_Delta_1}
Let $A$ be an NTWA.
Let a pattern $\Delta$ be a composition of any number of elements $\Delta_0$, $\Delta_1$, $\Delta_2$.
Then, $\Delta[*, \ldots, *]$ is equivalent to all patterns
obtained by attaching $\Delta_1$ to any port:
$\Delta_1(\Delta[*, \ldots, *])$,
$\Delta[*, \ldots, *, \Delta_1[*], *, \ldots, *]$.
\end{lemma}

\begin{figure}[t]
	\centerline{\includegraphics[scale=1]{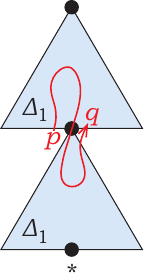}}
	\caption{Inner loop $p \to_{\epsilon} q$.}
	\label{f:inner_loop}
\end{figure}

Consider a computation of the automaton $A$ on a pattern $\Delta$
composed of elements $\Delta_1$ and $\Delta_2$.
This computation naturally splits into runs over the constituent elements.
The automaton possibly returns to the same ports of some elements multiple times,
and this complicates the analysis of such a computation.
In order to handle the returns to the same ports,
Boja\'nczyk and Colcombet~{\cite[\S 3.2]{BojanczykColcombet_reg}}
introduced the notion of \emph{inner loop}.
An inner loop from a state $p$ to a state $q$
is a computation in the pattern $\Delta_1[\Delta_1[*]]$
starting in the state $p$ at the junction node between two elements $\Delta_1$,
and ending in the state $q$ at the same junction node,
without visiting either port of the pattern $\Delta_1[\Delta_1[*]]$ on the way
(see Figure~\ref{f:inner_loop}).
The existence of such an inner loop is denoted by $p \to_{\epsilon} q$.
The computation in this definition is allowed to be empty,
and hence $p \to_{\epsilon} p$ for all $p$.

Some computations in patterns that return to their point of origin
can be shrunk to just inner loops.
\begin{lemma}[Boja\'nczyk and Colcombet~{\cite[Lemma 3.3]{BojanczykColcombet_reg}}]
\label{lemma_inner_loops}
Let $A$ be an NTWA.
Let $\Delta,\Delta'$ be patterns of nonzero rank
obtained as compositions of any number of elements $\Delta_0$, $\Delta_1$, $\Delta_2$.
Let the pattern $\Delta'$ be attached to the $i$-th leaf port of $\Delta$.
Assume that the automaton $A$ begins at the junction node between $\Delta$ and $\Delta'$ in some state $p$,
moves over these patterns without visiting their ports except the junction node,
and returns to the junction node in some state $q$.
Then $p \to_{\epsilon} q$.
\end{lemma}

Every inner loop can be executed at the junction node
between any two neighbouring elements $\Delta_0$, $\Delta_1$, $\Delta_2$, $\Delta_a$.
Indeed, by Lemma~\ref{lemma_attaching_Delta_1}
the elements $\Delta_0$, $\Delta_1$, $\Delta_2$ are equivalent to themselves
extended by attaching $\Delta_1$ to all ports including the root port.
As for $\Delta_a$,
the pattern $\Delta_1[\Delta_a]$ is equivalent to $\Delta_a$,
because $\Delta_1[\Delta_a] = \Delta_1[\Delta_1[\Delta'_a]]$,
and $\Delta_1[\Delta_1[*]]$ is equivalent to $\Delta_1[*]$
by Lemma~\ref{lemma_attaching_Delta_1}.

Then, runs between ports of these elements
can be regarded as runs between junction nodes,
each containing a pair $\Delta_1[\Delta_1[*]]$ in the vicinity,
and hence one can consider \emph{runs extended with loops},
which begin with an inner loop at the port of departure,
then make an ordinary run,
and finally execute another inner loop at the destination port.
For brevity, such runs shall be called \emph{transfers}.

\begin{definition}[Boja\'nczyk and Colcombet~{\cite[Defn.~5]{BojanczykColcombet_reg}}]
Let $A$ be an NTWA with a set of states $Q$.
Let $\Delta$ be a pattern of rank $k$,
let $p,q \in Q$ and $i,j \in \{0, 1, \ldots, k\}$.
A \emph{transfer of type $(p,i,q,j)$} over $\Delta$
is a computation that begins with an inner loop $p \to_{\epsilon}p'$,
continues with a run of type $(p',i,q',j) \in \delta_{\Delta}$,
and ends with another inner loop $q' \to_{\epsilon} q$,
where $p', q' \in Q$ are some states.

The \emph{relation of transfers} over $\Delta$
is $\gamma_{\Delta} \subseteq Q \times \{0, 1, \ldots, k\} \times Q \times \{0, 1, \ldots, k\}$,
and it contains the types of all transfers of $A$ over $\Delta$.
\end{definition}

If two patterns $\Delta$ and $\Delta'$,
composed of $\Delta_0$, $\Delta_1$, $\Delta_2$ and $\Delta_a$, are equivalent,
then their relations $\gamma_{\Delta}$ and $\gamma_{\Delta'}$ coincide:
indeed, the definition of a transfer depends only on the relation $\delta$,
whereas inner loops can be executed at any junction nodes.

A run through a pattern made of elements
naturally splits into runs through these elements.
Such a partition also can be made for transfers, as follows. 
\begin{definition}
Let a pattern $\Delta$ be obtained as a composition
of any number of elements $\Delta_1$ and $\Delta_2$.
For every transfer over $\Delta$,
its \emph{partition into elementary transfers}
over constituent elements $\Delta_1$ and $\Delta_2$
is obtained by first splitting this transfer
into inner loops and runs between neighbouring junction nodes,
and then attaching every inner loop to the preceding run
to obtain a transfer crossing this element
(inner loops at the beginning are attached to the following run).

A transfer is called \emph{simple}
if, in the above partition,
at most one elementary transfer crosses each element.
\end{definition}

This notion of a simple transfer
is analogous to a simple path in a tree.

\begin{lemma}\label{lemma_gamma_simple_path}
Let $A$ be an NTWA,
and let $\Delta$ be a pattern obtained by a composition of any number of elements $\Delta_1$ and $\Delta_2$.
Then, for every transfer over $\Delta$
there is a simple transfer of the same type.
\end{lemma}
\begin{proof}
Consider a transfer of type $(p,i,q,j)$ on the pattern $\Delta$,
and its partition into elementary transfers.

If at least two elementary transfers in this partition
end in the same junction node,
then, by Lemma~\ref{lemma_inner_loops},
the entire computation between these visits to the junction node
can be replaced with an inner loop.
Then one can
attach the resulting inner loop to a neighbouring transfer.
After such replacements we obtain a transfer of the same type $(p,i,q,j)$
in which all elementary transfers end in different junction nodes.

Now assume that the partition still contains multiple transfers through some element.
For every transfer, the next transfer 
through this element must begin at the same junction node
in which the previous transfer ends.
Then, among the multiple transfers through an element,
the first and the second ones must be consecutive in the partition,
for otherwise there would be two transfers ending in the same junction node: 
the first and the one before the second.

\begin{figure}[t]
	\centerline{\includegraphics[scale=1]{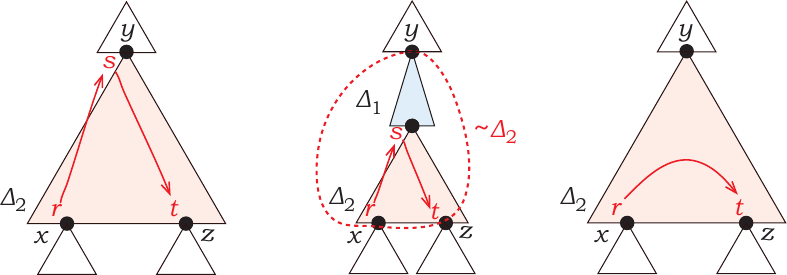}}
	\caption{Simplifying a transfer:
		(left) two consecutive transfers through the same element;
		(middle) inserting $\Delta_1$ at port $y$;
		(right) the resulting transfer of type $(r, x, t, z)$.}
	\label{f:simple_gamma_transfer}
\end{figure}

Let $(r,x,s,y)$ and $(s,y,t,z)$ be the types of these two consecutive transfers
through the same element $\Delta'$,
shown in Figure~\ref{f:simple_gamma_transfer}(left), where $\Delta'=\Delta_2$.
By Lemma~\ref{lemma_attaching_Delta_1},
this element $\Delta'$ is equivalent to the element $\Delta''$
obtained by attaching $\Delta_1$ to port $y$.
In $\Delta''$, there is a transfer of type $(r, x, t, z)$
obtained by first applying a transfer of type $(r,x,s,y)$ and then of type $(s,y,t,z)$,
both in the subelement $\Delta'$ of the element $\Delta''$,
as in Figure~\ref{f:simple_gamma_transfer}(middle).
And because $\Delta'$ and $\Delta''$ are equivalent,
a transfer of the same type is also possible in $\Delta'$,
see Figure~\ref{f:simple_gamma_transfer}(right).
Then, in the partition,
one can replace two transfers through $\Delta'$ with a transfer of type $(r, x, t, z)$.

After making all such replacements,
the desired simple transfer is obtained.
\end{proof}

\begin{figure}[t]
	\centerline{\includegraphics[scale=1]{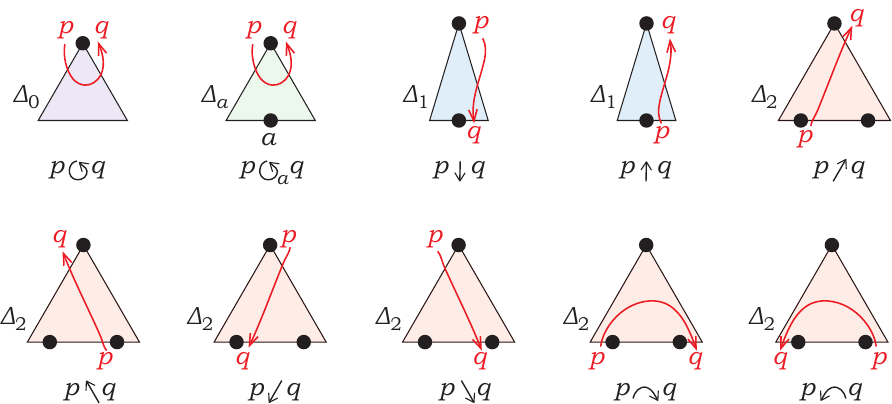}}
	\caption{Notation for transfers through elements $\Delta_0$, $\Delta_1$, $\Delta_2$ and $\Delta_a$.}
	\label{f:gamma_transfers_over_elements}
\end{figure}

The following notation for transfers
of the automaton $A$ through elements
$\Delta_0$, $\Delta_1$, $\Delta_2$ and $\Delta_a$
is introduced
(see Boja\'nczyk and Colcombet~\cite[Fig.~5.1]{BojanczykColcombet_reg}).
\begin{align*}
p \circlearrowleft q && \text{if } (p, 0, q, 0) \in \gamma_{\Delta_0}\\
p \circlearrowleft_a q && \text{if } (p, 0, q, 0) \in \gamma_{\Delta_a}\\
p \downarrow q && \text{if } (p, 0, q, 1) \in \gamma_{\Delta_1}\\
p \uparrow q && \text{if } (p, 1, q, 0) \in \gamma_{\Delta_1}\\
p \nearrow q && \text{if } (p, 1, q, 0) \in \gamma_{\Delta_2}\\
p \nwarrow q && \text{if } (p, 2, q, 0) \in \gamma_{\Delta_2}\\
p \swarrow q && \text{if } (p, 0, q, 1) \in \gamma_{\Delta_2}\\
p \searrow q && \text{if } (p, 0, q, 2) \in \gamma_{\Delta_2}\\
p \curvearrowleft q && \text{if } (p, 2, q, 1) \in \gamma_{\Delta_2}\\
p \curvearrowright q && \text{if } (p, 1, q, 2) \in \gamma_{\Delta_2}
\end{align*}
This notation is illustrated in Figure~\ref{f:gamma_transfers_over_elements}.
Note that the notation $p \diamond q$, with $\diamond \in \{\circlearrowleft, \ldots, \curvearrowright\}$,
always refers to a transfer from the state $p$ to the state $q$,
regardless of the direction of the arrow.
For example, $p \curvearrowleft q$ denotes that there is a transfer
from $p$ at port $2$ to $q$ at port $1$ in the element $\Delta_2$.

By Lemma~\ref{lemma_attaching_Delta_1}, attaching the element $\Delta_1$
to any port of $\Delta_2$ or $\Delta_1$
produces an equivalent element.
Therefore, for any transfer continued through an attached element $\Delta_1$,
there is a transfer of the same type through the original element.
The following lemma states a few such cases that are used in the paper.

\begin{lemma} \label{lemma_swallowing_delta_1}
Let $A$ be an NTWA with a set of states $Q$.
Then, for all states $p,q,r \in Q$,
\begin{enumerate}\renewcommand{\theenumi}{\alph{enumi}}
	\renewcommand{\labelenumi}{(\alph{enumi})}
\item	\label{lemma_swallowing_delta_1__p_up_q_up_r__p_up_r}
	$p \uparrow q \uparrow r$ implies $p \uparrow r$,
\item	\label{lemma_swallowing_delta_1__p_ne_q_up_r__p_ne_r}
	$p \nearrow q \uparrow r$ implies $p \nearrow r$,
\item	\label{lemma_swallowing_delta_1__p_up_q_ne_r__p_ne_r}
	$p \uparrow q \nearrow r$ implies $p \nearrow r$,
\item	\label{lemma_swallowing_delta_1__p_nw_q_up_r__p_nw_r}
	$p \nwarrow q \uparrow r$ implies $p \nwarrow r$,
\item	\label{lemma_swallowing_delta_1__p_up_q_right_r__p_right_r}
	$p \uparrow q \curvearrowright r$ implies $p \curvearrowright r$.
\end{enumerate}
\end{lemma}

The converse is also true:
if there is a transfer of type $(p,i,r,j)$ in the element $\Delta_1$
or $\Delta_2$ that moves from port $i$ to port $j$,
then a transfer of this type is also possible on an equivalent element
obtained by attaching $\Delta_1$ to port $i$ or $j$.
The latter transfer passes through the junction node,
and can be split into two transfers.

\begin{lemma} \label{lemma_inserting_delta_1}
Let $A$ be an NTWA, let $Q$ be its set of states.
Then, for all states $p,r \in Q$,
\begin{enumerate}\renewcommand{\theenumi}{\alph{enumi}}
	\renewcommand{\labelenumi}{(\alph{enumi})}
\item \label{lemma_inserting_delta_1_p_nearrow_q}
	$p \nearrow r$ implies $p \nearrow q \uparrow r$, for some $q \in Q$;
\item \label{lemma_inserting_delta_1_p_nwarrow_q}	
	$p \nwarrow r$ implies $p \nwarrow q \uparrow r$, for some $q \in Q$;
\item \label{lemma_inserting_delta_1_p_curvearrowright_q}	
$p \curvearrowright r$ implies $p\curvearrowright q \downarrow r$, for some $q \in Q$.
\end{enumerate}
\end{lemma}
\begin{proof}
By Lemma~\ref{lemma_gamma_simple_path},
for every transfer of type $(p,i,r,j)$ through an element extended with $\Delta_1$,
there is a simple transfer of this type through the extended element.
This transfer, by definition, can be split into two transfers,
one through the original element, and the other through the attached $\Delta_1$.
\end{proof}

Another basic lemma
says that all transfers between any leaf ports of patterns
made of $\Delta_0$, $\Delta_1$ and $\Delta_2$
can be reproduced on the element $\Delta_2$.

\begin{lemma}\label{lemma_from_big_elements_to_Delta_2}
Let $A$ be an NTWA with a set of states $Q$.
Let $\Delta$ be a $k$-ary pattern, with $k \geqslant 2$, 
constructed by composition of any number of elements $\Delta_0$, $\Delta_1$ and $\Delta_2$.
Let $(p,i,q,j) \in \gamma_\Delta$, for some states $p,q \in Q$ and
for some port numbers $i$ and $j$, with $1 \leqslant i,j \leqslant k$ and $i \neq j$.
Then, if $i < j$, then $p \curvearrowright q$,
and if $i > j$, then $p \curvearrowleft q$.
\end{lemma}
\begin{proof}
Consider any transfer of type $(p,i,q,j)$.
This computation does not visit any ports of the pattern $\Delta$ except $i$ and $j$.
Hence, one can attach elements $\Delta_0$
to all leaf ports except $i$ and $j$,
obtaining a new pattern $\Delta'$,
without affecting the computation.
This new pattern $\Delta'$ has two leaf ports,
so by Lemma~\ref{lemma_delta_123} it is equivalent to $\Delta_2$.

If $i < j$, then port $i$ becomes port $1$, and port $j$ becomes port $2$,
and the original transfer on the pattern $\Delta$
turns into a transfer of type $(p,1,q,2)$
on a pattern equivalent to $\Delta_2$.
Thus, $p \curvearrowright q$.

Similarly, if $i > j$, then port $i$ becomes port $2$, and port $j$ becomes port $1$.
In this case, a transfer of type $(p,2,q,1)$ is obtained, that is, $p \curvearrowleft q$.
\end{proof}

The next lemma asserts that,
as long as an automaton can move from the left leaf port of $\Delta_2$
to its right leaf port,
starting in a state $p$ and ending in a state $q$,
it can either similarly move
in every pattern constructed from elements $\Delta_0$, $\Delta_1$ and $\Delta_2$
from an arbitrary leaf port to the next leaf port in order,
or there is an inner loop from state $p$ to state $q$.

\begin{lemma}\label{lemma_transfers_through_Delta2_in_big_elements}
Let $A$ be an NTWA with a set of states $Q$.
Let $\Delta$ be a pattern of rank $k$, with $k \geqslant 2$, 
made of any number of elements $\Delta_0$, $\Delta_1$ and $\Delta_2$.
Let $p,q \in Q$ be some states with $p \curvearrowright q$.
Then either $(p,i,q,i+1)\in \gamma_\Delta$ for all $i$ with $1 \leqslant i \leqslant k-1$,
or $p \to_\epsilon q$.
\end{lemma}
\begin{proof}
For every number $i$, with $1 \leqslant i \leqslant k-1$, it should be proved
that either $(p,i,q,i+1)\in \gamma_\Delta$, or $p \to_\epsilon q$.
Let the number $i$ be fixed.

First, let $\Delta_{i,i+1}$ be the pattern obtained from the pattern $\Delta$
by attaching elements $\Delta_0$ to all leaf ports except $i$ and $i+1$.
The resulting pattern is equivalent to $\Delta_2$
by Lemma~\ref{lemma_delta_123}.
Since $p \curvearrowright q$,
there is a transfer on the pattern $\Delta_{i,i+1}$,
from the state $p$ in port $1$ (formerly, port $i$)
to the state $q$ in port $2$ (formerly, port $i+1$).
If this transfer never visits any of the attached elements $\Delta_0$,
then $(p,i,q,i+1)\in \gamma_\Delta$, as desired.
Now assume that this transfer visits some elements $\Delta_0$,
and let the last visit to any $\Delta_0$
happen in some state $p_1$ in some port $j$ of the original pattern $\Delta$,
where $1 \leqslant j \leqslant k$, $j \neq i$ and $j \neq i+1$.
This computation is illustrated in Figure~\ref{f:transfers_through_Delta2_in_big_elements}(left).

Assume that $j<i$; the case of $j>i+1$ can be proved symmetrically.

\begin{figure}[t]
	\centerline{\includegraphics[scale=.9]{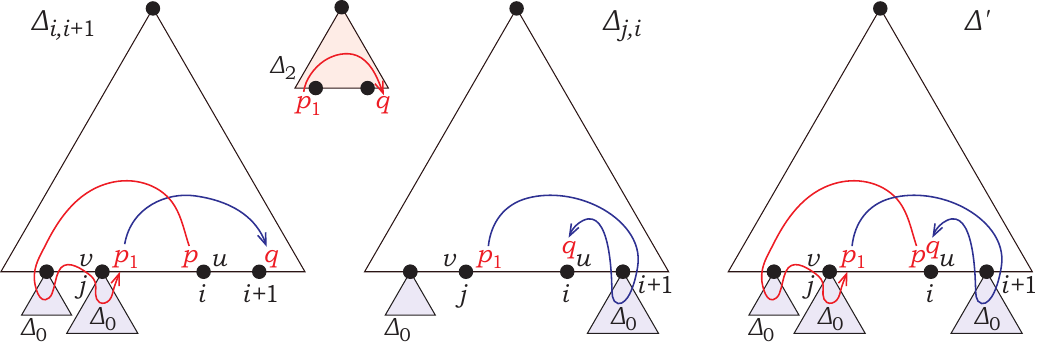}}
	\caption{(left) the computation on $\Delta_{i,i+1}$;
		(middle) the transfer of type $(p_1, 1, q, 2)$ on $\Delta_{j,i}$;
		(right) the two computations combined on the pattern $\Delta'$.}
	\label{f:transfers_through_Delta2_in_big_elements}
\end{figure}

Denote by $u$ in all patterns the node which is port $i$ in the pattern $\Delta$,
and denote by $v$ the node which is port $j$ in $\Delta$.
The existing transfer of type $(p,1,q,2)$ through the pattern $\Delta_{i,i+1}$
can be split into two parts:
up to the last visit to the node $v$ in the state $p_1$,
and thereafter.
The second part of this computation moves through $\Delta_{i,i+1}$
without visiting any attached elements $\Delta_0$,
and hence it actually proceeds through the pattern $\Delta$.
This part of the computation is a transfer of type $(p_1,j,q,i+1)$
through the original pattern $\Delta$.
Then, by Lemma~\ref{lemma_from_big_elements_to_Delta_2},
$p_1 \curvearrowright q$.

Let $\Delta_{j,i}$ be the pattern obtained from $\Delta$
by attaching elements $\Delta_0$ to all ports except $i$ and $j$,
as in Figure~\ref{f:transfers_through_Delta2_in_big_elements}(middle).
Then $\Delta_{j,i}$ is equivalent to $\Delta_2$ by Lemma~\ref{lemma_delta_123},
and since $p_1 \curvearrowright q$,
there is a transfer in $\Delta_{j,i}$
from configuration $(p_1, v)$ to configuration $(q, u)$.

Finally, $\Delta'=\Delta_{i,i+1}[*, \Delta_0]=\Delta_{j,i}[\Delta_0, *]$
is the pattern obtained from $\Delta$
by attaching elements $\Delta_0$ to all ports except $i$,
as shown in Figure~\ref{f:transfers_through_Delta2_in_big_elements}(right).
Then a transfer of type $(p,1,q,1)$ in the pattern $\Delta'$
can be constructed by combining two computations:
first, the initial part of the computation on $\Delta_{i,i+1}$ that leads from $(p, u)$ to $(p_1, v)$,
and then a transfer on the pattern $\Delta_{j,i}$ that starts in $(p_1, v)$ and ends in $(q, u)$.
The resulting transfer is the desired inner loop
by Lemma~\ref{lemma_inner_loops}
(to use this lemma, one can attach an element $\Delta_1$ to port $1$ of $\Delta'$).
\end{proof}

A convenient tool used by Boja\'nczyk and Colcombet~\cite[Sect.~2.1]{BojanczykColcombet_reg}
is an assumption that
a tree-walking automaton is \emph{time-symmetric},
in the sense that for every computation it can make,
it can also make a related computation
that proceeds in the reverse direction.
For a time-symmetric automaton, the number of cases in many proofs can be halved.
Furthermore, every nondeterministic tree-walking automaton can be made trivially time-symmetric
by adding unreachable states with reversed transitions.
This is stated in the following lemma.

\begin{lemma}\label{symmetries_lemma}
For every NTWA $B=(\Sigma, Q, Q_0, \delta, F)$
there exists another NTWA $\widehat{B}=(\Sigma, \widehat{Q}, Q_0, \widehat{\delta}, F)$,
with $|\widehat{Q}|$ even,
which recognizes the same set of trees as $B$,
and is \emph{time-symmetric}, in the sense that
there exists a bijection $\tau \colon \widehat{Q} \to \widehat{Q}$,
such that for every pattern $\Delta$,
for every two ports $i, j$ in the pattern,
and for every two states $\widehat{p}, \widehat{q} \in \widehat{Q}$,
the following two statements are equivalent.
\begin{enumerate}
\item
	There is a run
	of type $(\widehat{p}, i, \widehat{q}, j)$ of $\widehat{B}$ on $\Delta$.
\item
	There is a run
	of type $(\tau(\widehat{q}), j, \tau(\widehat{p}), i)$ of $\widehat{B}$ on $\Delta$.
\end{enumerate}
Furthermore, the same equivalence holds for transfers over $\Delta$.
\end{lemma}
\begin{proof}
The set of states of the desired NTWA is $\widehat{Q}=Q \cup Q'$,
where $Q'=\set{q'}{q \in Q}$,
and the bijection $\tau$ translates between $Q$ and $Q'$
as $\tau(q)=q'$ and $\tau(q')=q$ for all $q \in Q$.
For every transition 
	$(p, a, \mathrm{type}_1,
	q, b, \mathrm{type}_2,
	d) \in \delta$,
the new automaton has this transition and also the following reversed transition on new states.
\begin{equation*}
	(q', b, \mathrm{type}_2, p', a, \mathrm{type}_1, -d)
\end{equation*}
Thus, there are no transitions between subsets of states $Q$ and $Q'$.

Assume that there is a run of type $(\widehat{p}, i, \widehat{q}, j)$
on some pattern $\Delta$.
Then, either $\widehat{p},\widehat{q} \in Q$ or $\widehat{p},\widehat{q} \in Q'$.
If both states are in $Q$, let $\widehat{p}=p$ and $\widehat{q}=q$.
Then the run begins in state $p$ in port $i$
and ends in state $q$ in port $j$, with all intermediate states in $Q$.
Then, $\tau$ maps each state $r$ in this computation to $r'$,
and each transition has a reversed copy on $Q'$.
Those reversed transitions form a run from $q'$ in port $j$ to $p'$ in port $i$,
which is of the desired type $(q', j, p', i)$.

The case when both $\widehat{p}$ and $\widehat{q}$ are in $Q'$ is completely symmetric.

The converse implication in the lemma follows, because $\tau$ applied twice is the identity function.

The last claim about transfers follows from the equivalence for runs
by considering computations on $\Delta$ with two elements $\Delta_1$ attached to ports $i$ and $j$.
In particular, computations implementing inner loops are also reversed.
\end{proof}

The use of time symmetry shall now be illustrated
on the following lemma, which states that
if the automaton can move through $\Delta_1$ upwards,
then it can move upwards through $\Delta_2$ from at least one of the leaf ports using the same states;
and the same result holds for downward motion.
The proof of the upward case can be found
in the paper by Boja\'nczyk and Colcombet~\cite{BojanczykColcombet_reg},
and, for demonstration purposes, the downward case shall now be inferred from the upward case.

\begin{lemma}[Boja\'nczyk and Colcombet~{\cite[Prop.~5.6]{BojanczykColcombet_reg}}]
\label{lemma_delta1_up_and_down}
Let $A$ be a time-symmetric NTWA.
Then, for all states $p$ and $q$, $p \uparrow q$ if and only if $p \nearrow q$ or $p \nwarrow q$.
Similarly, $p \downarrow q$ if and only if $p \swarrow q$ or $p \searrow q$.
\end{lemma}
\begin{proof}[Proof of the second part]
For states $p$ and $q$, consider the corresponding states $\tau(p)$ and $\tau(q)$,
where $\tau$ is the bijection from Lemma~\ref{symmetries_lemma}.
Then, $p \downarrow q$ is equivalent to $\tau(q) \uparrow \tau(p)$,
and $p \swarrow q$ is equivalent to $\tau(q) \nearrow \tau(p)$,
and $p \searrow q$ is equivalent to $\tau(q) \nwarrow \tau(p)$. 
Then the second part of the lemma for states $p$ and $q$
follows from the first part of the lemma applied to states $\tau(q)$ and $\tau(p)$.
\end{proof}

\section{Patterns $\Delta_n$ and $\Delta_{2M}'$}\label{section_elements_with_and_without_errors}

Let $A$ be an NTWA, it should be proved that it does not recognize the complement of $L$.
Let $Q$ be its set of states and let $n=|Q|$.
In view of Lemma~\ref{symmetries_lemma},
it can be safely assumed that the automaton $A$
is time-symmetric, and $n$ is even and is at least $4$.
This automaton $A$ is fixed for the rest of the paper.

The goal of the proof is to construct two trees:
a tree not in $L$ and a modified tree in $L$, such that if the automaton $A$ accepts
the first tree, then it can be lured to accept the second tree.
These two trees are constructed in the form of two large patterns of rank $1$,
which are completed into trees by attaching a root with $\Delta_1$ at the top
and $\Delta_0$ at the bottom.

Let $M$ be a number with residue $\frac{n}{2}$ modulo $n!$,
such that $M > n^2+10n$.

\begin{figure}[t]
	\centerline{\includegraphics[scale=.9]{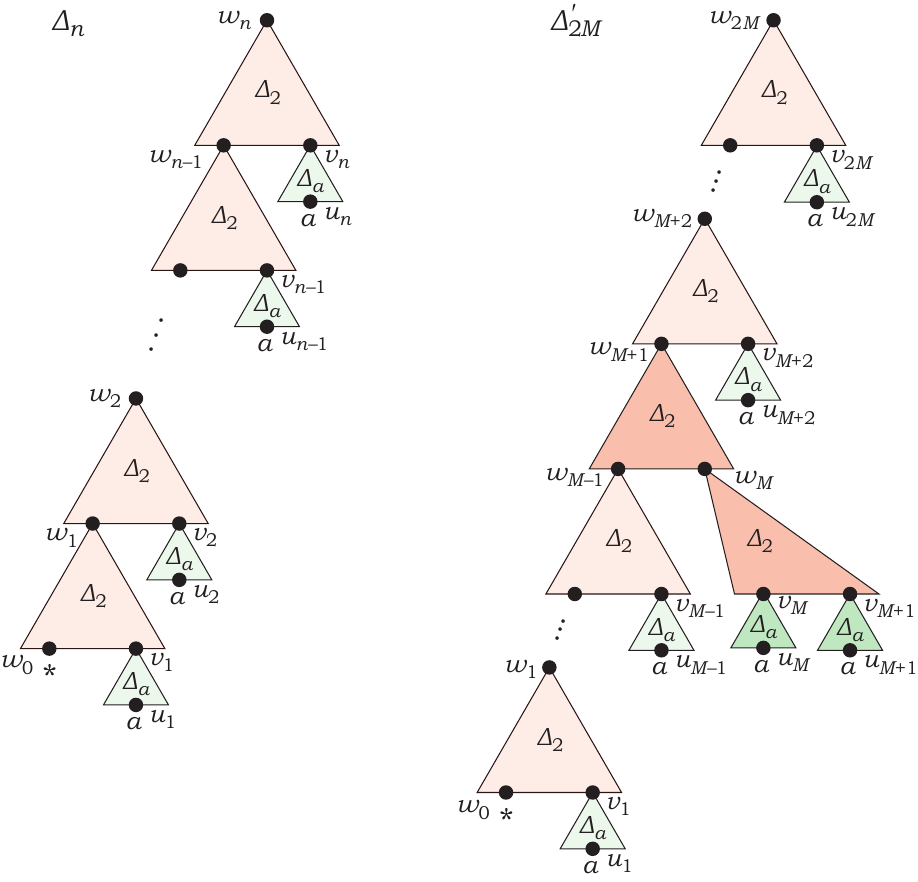}}
	\caption{Patterns $\Delta_n$ and $\Delta_{2M}'$.}
	\label{f:patterns_Delta_n_Delta_2M_prime}
\end{figure}

The desired patterns are denoted by $\Delta_{n}$ and $\Delta_{2M}'$
and are illustrated in Figure~\ref{f:patterns_Delta_n_Delta_2M_prime}.
The pattern $\Delta_n$, called \emph{the small correct pattern},
is a chain of $n$ elements $\Delta_2[*,\Delta_a]$ of rank $1$.
The other pattern $\Delta_{2M}'$, \emph{the faulty pattern},
is constructed by attaching to the element $\Delta_2[*,\Delta_2[\Delta_a,\Delta_a]]$
two chains of $M-1$ elements $\Delta_2[*,\Delta_a]$ each,
one to the leaf port and the other to the root port of the central element.
We shall call this central part \emph{the fault},
since the resulting tree will be in $L$
due to the different structure of $a$-leaves in this element.

In both patterns $\Delta_n$ and $\Delta_{2M}'$,
all leaves labelled with $a$ are enumerated from left to right;
denote these leaves by $u_1, \ldots, u_n$ in the small correct pattern,
and by $u_1, \ldots, u_{2M}$ in the faulty pattern.
For convenience, the root ports of elements $\Delta_a$ containing these leaves
are denoted by $v_1, \ldots, v_n$ in the small correct pattern
and by $v_1, \ldots, v_{2M}$ in the faulty pattern,
whereas the root ports of the elements $\Delta_2[*,\Delta_a]$,
are denoted by $w_1,\ldots,w_n$ in the small correct pattern
and $w_1,\ldots,w_{2M}$ in the faulty pattern.
Let the leaf port be $w_0$ in both patterns.
Note that, in the faulty pattern, $w_M$ is the root port of $\Delta_2[\Delta_a,\Delta_a]$,
and $w_{M+1}$ is the root port of $\Delta_2[*,\Delta_2[\Delta_a,\Delta_a]]$:
these two nodes are enumerated out of order of traversal.

In the small correct pattern $\Delta_n$, all triples of consecutive $a$-leaves 
$u_i$, $u_{i+1}$, $u_{i+2}$
do not satisfy $\mathrm{lca}(u_i, u_{i+1},u_{i+2}) = \mathrm{lca}(u_i,u_{i+1})$,
whereas in $\Delta_{2M}'$, there is a triple $u_{M-1}, u_{M}, u_{M+1}$
with $\mathrm{lca}(u_{M-1},u_{M}, u_{M+1}) = \mathrm{lca}(u_{M-1},u_M)$.
Therefore, if the small correct pattern $\Delta_n$ is replaced with $\Delta_{2M}'$
in the tree $\mathit{root}[\Delta_1[\Delta_n[\Delta_0]]]$ that is not in $L$,
then the resulting tree will be in $L$.
The goal is to prove that if the automaton $A$
accepts the tree $\mathit{root}[\Delta_1[\Delta_n[\Delta_0]]]$,
then it also accepts the tree $\mathit{root}[\Delta_1[\Delta_{2M}'[\Delta_0]]]$.
This will show that the automaton $A$ does not recognize the complement of $L$,
which is enough to prove Theorem~\ref{ntwa_rec_L_but_not_compl_of_L}.

Consider any accepting computation of the automaton $A$
on the tree $\mathit{root}[\Delta_1[\Delta_n[\Delta_0]]]$.
This computation is split into segments between visits to the ports of the small correct pattern $\Delta_n$.
Segments that lie outside $\Delta_n$
can be exactly replicated in the tree $\mathit{root}[\Delta_1[\Delta_{2M}'[\Delta_0]]]$.
The rest of the computation is formed by runs through the small correct pattern $\Delta_n$,
which have to be reproduced on the faulty pattern $\Delta_{2M}'$.
That is, for every run of type $(p,i,q,j) \in \delta_{\Delta_n}$
one should prove that $(p,i,q,j) \in \delta_{\Delta_{2M}'}$.
Then it is enough to prove the following lemma.

\begin{mainlemma}
Let $A$ be a time-symmetric nondeterministic tree-walking automaton,
with the set of states $Q$ of even size $n \geqslant 4$,
operating on binary trees with labels $\{a,b\}$,
let $\Delta_0$, $\Delta_1$, $\Delta_2$ and $\Delta_a$
be the elements constructed for this automaton
as in Section~\ref{section_methods_BojanczykColcombet},
and let the patterns $\Delta_n$ and $\Delta_{2M}'$ be as defined above.
Then $\delta_{\Delta_n} \subseteq \delta_{\Delta_{2M}'}$.
\end{mainlemma}

Note that the faulty pattern $\Delta_{2M}'$ begins and ends with two subpatterns $\Delta_n$,
that is, it can be represented as $\Delta_n[\Delta_{2M-2n}'[\Delta_n[*]]]$.
This implies that if the automaton can return to the same port in $\Delta_n$,
changing its state from $p$ to $q$, then it can do the same in $\Delta_{2M}'$.
That is, if $(p,i,q,i) \in \delta_{\Delta_n}$, for some states $p,q \in Q$ and some port $i \in \{0, 1\}$,
then $(p,i,q,i) \in \delta_{\Delta_{2M}'}$.

Now it remains to prove that for every run through the small correct pattern $\Delta_n$
from the leaf port to the root port (or from the root port to the leaf port),
there is a run of the same type
through the faulty pattern $\Delta_{2M}'$.
Since the automaton $A$ is time-symmetric,
the proof will be given only for runs from leaf to root,
that is, that $(p,1,q,0) \in \delta_{\Delta_n}$ implies $(p,1,q,0) \in \delta_{\Delta'_{2M}}$.
Then, for runs from root to leaf, if $(p,0,q,1) \in \delta_{\Delta_n}$,
then $(\tau(q),1,\tau(p),0) \in \delta_{\Delta_n}$,
which is a run from leaf to root, and its existence implies
$(\tau(q),1,\tau(p),0) \in \delta_{\Delta'_{2M}}$
by the case of an upward run.
Finally, by time symmetry again, $(p,0,q,1) \in \delta_{\Delta'_{2M}}$.

Some states $q_{\text{start}}$ and $q_{\text{finish}}$
with $(q_{\text{start}},1,q_{\text{finish}},0) \in \delta_{\Delta_n}$
are fixed for the rest of the proof.
And it should be proved that $(q_{\text{start}},1,q_{\text{finish}},0) \in \delta_{\Delta_{2M}'}$.

\section{The correct pattern $\Delta_{2M}$ and a run of type $(q_{\text{start}},1,q_{\text{finish}},0)$ through it}\label{section_passing_through_Delta_2M}

The ultimate goal is to construct a run of type $(q_{\text{start}},1,q_{\text{finish}},0)$
through the faulty pattern $\Delta_{2M}'$,
which is much larger than the small correct pattern $\Delta_n$.
In this section, a run of this type is constructed for another pattern:
\emph{the correct pattern} $\Delta_{2M}$, which is an inflated version of $\Delta_n$.
It is composed of $2M$ elements $\Delta_2[*,\Delta_a]$ forming a chain.
In the pattern $\Delta_{2M}$, there are nodes $u_i,v_i,w_i$, with $i \in \{1,\ldots,2M\}$,
and $w_0$, which are defined analogously to the nodes of $\Delta_n$.

\begin{figure}[hbtp]
	\centerline{\includegraphics[scale=.9]{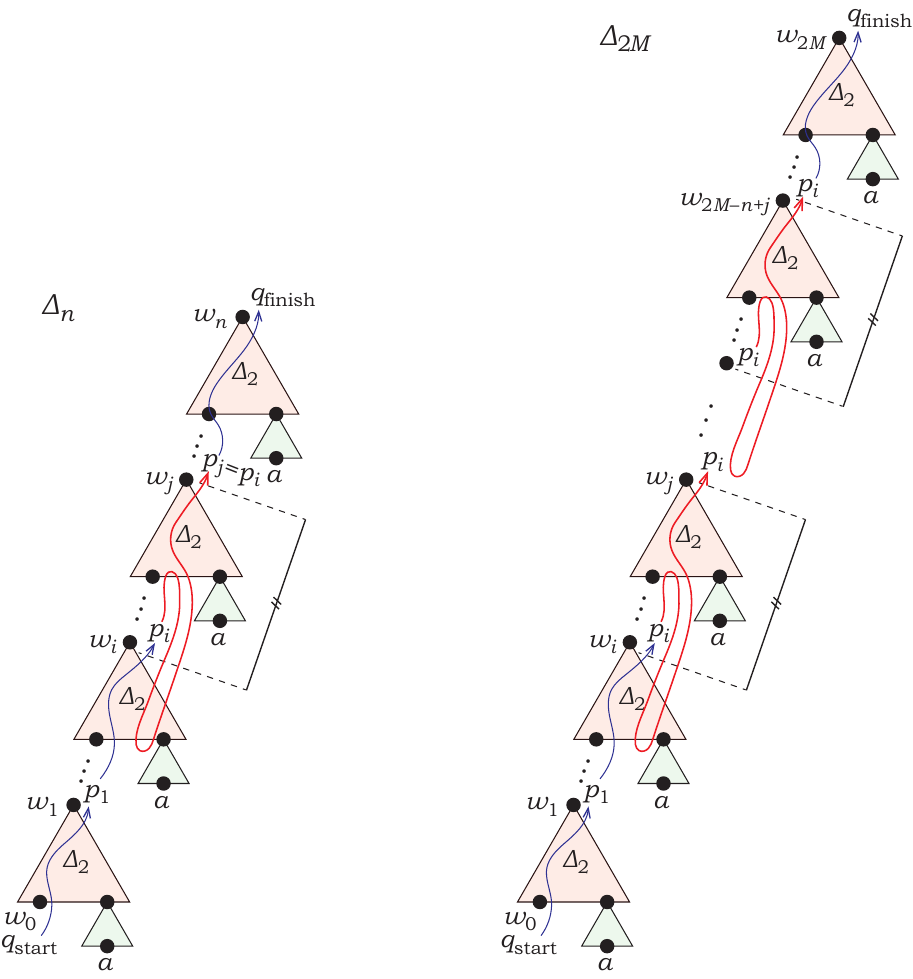}}
	\caption{(left) A run of type $(q_{\text{start}},1,q_{\text{finish}},0)$
		through $\Delta_n$, with its part from $(p_i, w_i)$ to $(p_i, w_j)$ to be used as a period;
		(right) A run of the same type through $\Delta_{2M}$, with multiple iterations of the period.}
	\label{f:run_through_Delta_2M}
\end{figure}

The first claim is that there is a run of type $(q_{\text{start}},1,q_{\text{finish}},0)$
through the correct pattern $\Delta_{2M}$,
which behaves periodically in the middle of the pattern,
and which is obtained by taking a run
through the small correct pattern $\Delta_n$
and repeating its part periodically,
as illustrated in Figure~\ref{f:run_through_Delta_2M}.

\begin{claim}\label{claim_run_Delta_2M_from_w_to_w}
There is a run of type $(q_{\text{start}},1,q_{\text{finish}},0)$
through the correct pattern $\Delta_{2M}$,
such that, for some $i$ with $0 \leqslant i \leqslant n-1$, after the first visit to the node $w_i$
the automaton starts periodically repeating a certain sequence of transitions,
with the prefix before periodic part and the first iteration together
confined to the bottom $n$ elements $\Delta_2[*,\Delta_a]$.
This periodic behaviour leads it to a node with number greater than $2M-n$,
and the last iteration together with the suffix after the periodic part
are confined to the top $n$ elements $\Delta_2[*,\Delta_a]$ of $\Delta_{2M}$.
\end{claim}
\begin{proof}
Consider the states $p_0, p_1, \ldots, p_n$, 
in which the automaton $A$ first visits the leaf port $w_0$ of $\Delta_n$
and the root ports $w_1,\ldots,w_n$ of elements $\Delta_2$ in $\Delta_n$
while making a run of type $(q_{\text{start}},1,q_{\text{finish}},0)$.
Here $p_0 = q_{\text{start}}$ and $p_n = q_{\text{finish}}$.
Among these states there are two equal states:
$p_i = p_j$, for some $i$ and $j$ with $0 \leqslant i < j \leqslant n$,
as shown in Figure~\ref{f:run_through_Delta_2M}(left).

The pattern $\Delta_{2M}$ begins and ends with the small correct pattern $\Delta_n$.
The desired run through $\Delta_{2M}$ is constructed
as in Figure~\ref{f:run_through_Delta_2M}(right):
it starts in the leaf port in the state $q_{\text{start}}$,
then, as in $\Delta_n$, it comes in the state $p_i$ to the node $w_i$,
which is the root port of the $i$-th element $\Delta_2[*,\Delta_a]$ from the bottom
(or the leaf port of $\Delta_{2M}$ if $i = 0$).
A sequence of transitions leading from configuration $(p_i,w_i)$ to configuration $(p_i,w_j)$
is taken from the computation of the automaton $A$ on $\Delta_n$.
On $\Delta_n$, this sequence never visits the leaf port on the way
and nevers moves up from $w_j$.
This sequence of transitions can be executed on the pattern $\Delta_{2M}$ 
from any configuration $(p_i, w_t)$, with $i \leqslant t \leqslant 2M-j+i$,
and the automaton will come to the configuration $(p_i,w_{t+j-i})$. 
This sequence visits at most $i$ different elements $\Delta_2[*,\Delta_a]$ below the starting node $w_t$, 
and exactly $j-i$ such elements above. 

Then the run through $\Delta_{2M}$
can be continued from the configuration $(p_i,w_i)$
by repeating this sequence of transitions periodically,
until the automaton comes to the configuration $(p_i, w_{2M-n+j})$.
It takes $\frac{2M-n}{j-i}+1$ repetitions to do so:
indeed, $2M-n$ is divisible by $j-i$,
because $2M$ gives residue $n$ modulo $n!$, and $j-i \leqslant n$,
and after $\frac{2M-n}{j-i}+1$ repetitions the automaton moves from the node $w_i$
to the node with number $i+\big(\frac{2M-n}{j-i}+1\big) \cdot (j-i) = i + 2M - n + j - i = 2M-n+j$.

Afterwards, the automaton
applies the transitions from its run on $\Delta_n$ as follows.
All the transitions it made on $\Delta_n$ starting from the 
moment of its first visit to the node $w_j$ in the state $p_i$
and ending in the root port in the state $q_{\text{finish}}$
are made on $\Delta_{2M}$ from the configuration $(p_i,w_{2M-n+j})$,
and the automaton similarly finishes
in the root port of $\Delta_{2M}$ in the state $q_{\text{finish}}$.

The first iteration of the repeated sequence together with the prefix
are taken from the computation on $\Delta_n$,
and hence are contained in the first $n$ elements $\Delta_2[*, \Delta_a]$.
Analogously, the last iteration together with the suffix
are also part of the computation on $\Delta_n$,
and therefore they fit in the last $n$ elements $\Delta_2[*, \Delta_a]$.
\end{proof}

The goal is to take the computation on the correct pattern $\Delta_{2M}$
from Claim~\ref{claim_run_Delta_2M_from_w_to_w},
and to reproduce it on the faulty pattern $\Delta_{2M}'$,
so that the periodically repeated sequence will pass by the fault,
that is, the element $\Delta_2[*,\Delta_2[\Delta_a,\Delta_a]]$,
paying no attention to the difference.

In order to distinguish the faulty pattern from the correct pattern,
the periodically repeated sequence
should visit some of the elements $\Delta_a$ attached from the right.
In the next claim it is shown that if none of the elements $\Delta_a$ are visited,
then the computation can be reproduced on the faulty pattern,
and in this case the proof of the Main Lemma will be completed.

\begin{claim}\label{claim_run_Delta_2M_from_w_to_w_never_visiting_Delta_a}
Consider some run of type $(q_{\text{start}},1,q_{\text{finish}},0)$
through the correct pattern $\Delta_{2M}$,
as constructed in Claim~\ref{claim_run_Delta_2M_from_w_to_w}.
Then, if its periodic part never visits any element $\Delta_a$,
there is a run of type $(q_{\text{start}},1,q_{\text{finish}},0)$
through the faulty pattern $\Delta'_{2M}$.
\end{claim}

\begin{figure}[t]
	\centerline{\includegraphics[scale=0.9]{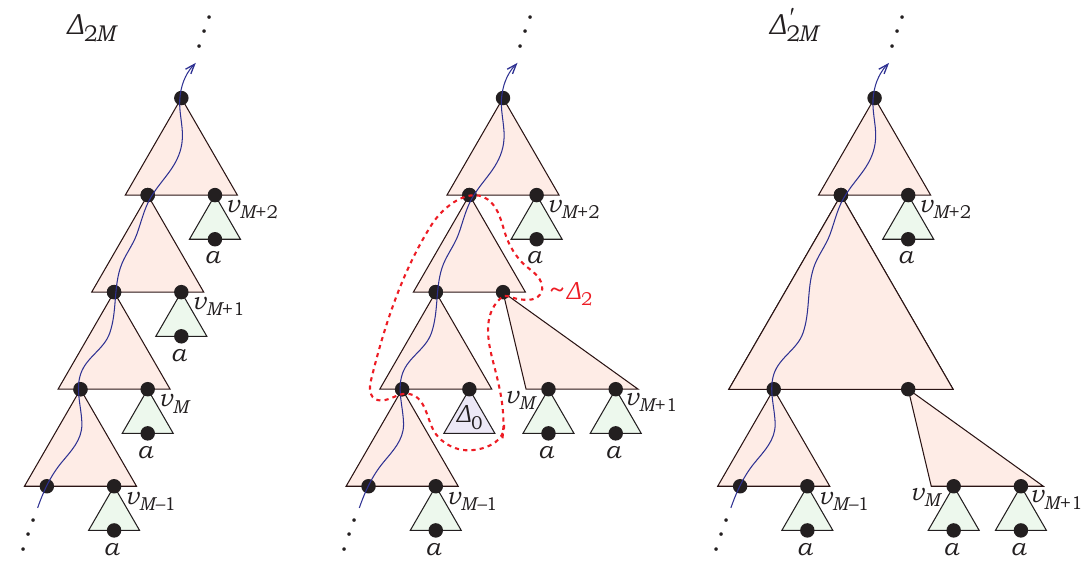}}
	\caption{(left) Computation on $\Delta_{2M}$ never visiting any $\Delta_a$;
		(middle) Two of $\Delta_a$ replaced with $\Delta_0$ and $\Delta_2[\Delta_a, \Delta_a]$;
		(right) An element replaced with an equivalent one, resulting with $\Delta'_{2M}$.}
	\label{f:if_repeated_sequence_never_visits_Delta_a}
\end{figure}

\begin{proof}
Assume that the periodically repeated part of the computation on $\Delta_{2M}$ never visits $\Delta_a$,
as shown in Figure~\ref{f:if_repeated_sequence_never_visits_Delta_a}(left).
Then one can replace any elements $\Delta_a$ from the $n$-th to the $(2M-n)$-th
with any elements of rank zero
without affecting the computation,
so that a run of type $(q_{\text{start}},1,q_{\text{finish}},0)$ will still exist.
This way, the subtree of the node $v_M$ is replaced with $\Delta_0$,
and the subtree of $v_{M+1}$ is replaced with $\Delta_2[\Delta_a,\Delta_a]$,
see Figure~\ref{f:if_repeated_sequence_never_visits_Delta_a}(middle).
As a result, the element $\Delta_2[\Delta_2[*,\Delta_a],\Delta_a]$ in $\Delta_{2M}$
is replaced with $\Delta_2[\Delta_2[*,\Delta_0],\Delta_2[\Delta_a,\Delta_a]]$.
By Lemma~\ref{lemma_delta_123}, the element $\Delta_2[\Delta_2[*,\Delta_0],*]$
encircled in Figure~\ref{f:if_repeated_sequence_never_visits_Delta_a}(middle) by a dotted line
is equivalent to the element $\Delta_2$.
After replacing $\Delta_2[\Delta_2[*,\Delta_0],*]$ with $\Delta_2$,
one obtains the faulty pattern $\Delta'_{2M}$,
as in Figure~\ref{f:if_repeated_sequence_never_visits_Delta_a}(right).
And since the replacement of elements with equivalent ones
does not change the function $\delta$,
it turns out that $(q_{\text{start}},1,q_{\text{finish}},0) \in \delta_{\Delta_{2M}'}$,
as desired.
\end{proof}

If no elements $\Delta_a$ are visited in the periodic part of the computation, 
then the proof of the Main Lemma is completed 
by Claim~\ref{claim_run_Delta_2M_from_w_to_w_never_visiting_Delta_a}. 
From now on, it is assumed that some elements $\Delta_a$ are visited in the periodic part.
The plan is to represent this periodic part
as a sequence of segments between visits to $\Delta_a$.
It will be proved that such segments
can be executed by simple transfers.
These segments shall be called \emph{proper steps}.

\begin{figure}[t]
	\centerline{\includegraphics[scale=0.9]{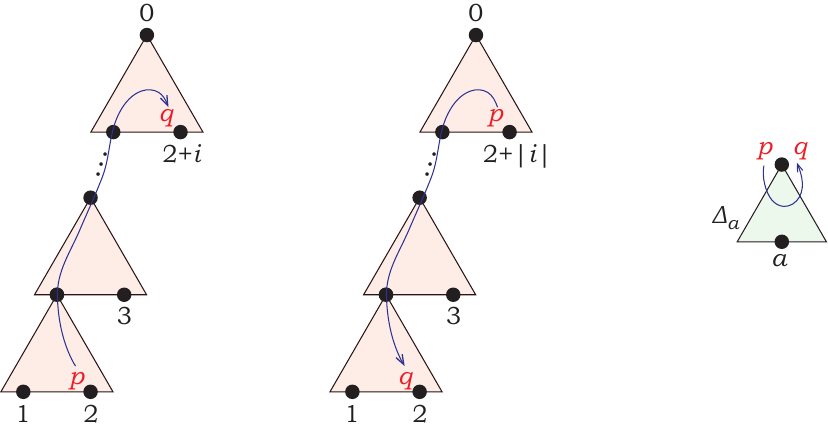}}
	\caption{A proper step of type $(i, p, q)$:
		(left) $i>0$; (middle) $i<0$; (right) $i=0$.}
	\label{f:step_definition}
\end{figure}

\begin{definition}\label{definition_step}
Let $i$ be an integer 
such that $i \neq 0$,
and let $p,q \in Q$ be two states.
Consider the pattern obtained by attaching $|i|+1$ elements $\Delta_2$ into a chain,
with every next element attached to the left leaf port of the previous one,
as illustrated in Figure~\ref{f:step_definition}.
Then a \emph{proper step of type $(i,p,q)$}
is a simple transfer through this pattern,
of type $(p, 2, q, 2+i)$ if $i>0$,
or of type $(p, 2+|i|, q, 2)$ if $i<0$.

For $i=0$ a proper step of type $(0,p,q)$
is a transfer of type $(p, 0, q, 0)$ on $\Delta_a$.

The number $i$ is called the \emph{pace} of the proper step.
\end{definition}

This definition can be conveniently reformulated
in the notation for transfers over elements $\Delta_2$.

\begin{claim}\label{claim_sequence_for_step_partition}
Let $p,q \in Q$ be two states, and let $i$ be a non-zero integer.
If $i > 0$, then a proper step of type $(i,p,q)$
exists if and only if
there are intermediate states $p_1,\ldots,p_i$,
such that $p \nwarrow p_1 \nearrow p_2 \nearrow \ldots \nearrow p_i \curvearrowright q$.
If $i < 0$, then a proper step of type $(i,p,q)$
exists if and only if
$p \curvearrowleft p_1 \swarrow p_2 \swarrow \ldots \swarrow p_i \searrow q$,
for some intermediate states $p_1,\ldots,p_i$.
\end{claim}
\begin{proof}
The proof is given for the case of $i>0$,
the other case of $i<0$ is analogous.
A proper step is a simple transfer through a chain of elements $\Delta_2$,
see Figure~\ref{f:step_definition}(left).
Assume that there is a proper step of type $(i,p,q)$.
This proper step splits into transfers over elements $\Delta_2$,
with some intermediate states $p_1, \ldots, p_i$,
and the transfers in this partition
are of types
$(p,2,p_1,0)$, $(p_1,1,p_2,0)$, \ldots, $(p_{i-1},1,p_i,0)$, $(p_i,1,q,2)$
in the order of the computation.
The existence of transfers of these types
is expressed by the formula
$p \nwarrow p_1 \nearrow p_2 \nearrow \ldots \nearrow p_i \curvearrowright q$.

Conversely, let the formula be true.
Then one can take any transfers
of types $(p,2,p_1,0)$, $(p_1,1,p_2,0)$, \ldots, $(p_{i-1},1,p_i,0)$, $(p_i,1,q,2)$,
and use them sequentially to obtain a proper step of type $(i,p,q)$.
\end{proof}

A proper step of type $(i,p,q)$ can be used anywhere in the pattern $\Delta_{2M}$:
that is, for all integers $x$ satisfying $1 \leqslant x \leqslant 2M$
and $1 \leqslant x+i \leqslant 2M$,
there is a sequence of transfers
on elements $\Delta_2$, that moves the automaton on the pattern $\Delta_{2M}$
from the configuration $(p, v_x)$ 
to the configuration $(q,v_{x+i})$
without visiting root ports of any $\Delta_a$ on the way,
and never traversing any $\Delta_2$ twice.

Next, a run through $\Delta_{2M}$ constructed in Claim~\ref{claim_run_Delta_2M_from_w_to_w}
is modified so that its periodic part consists of proper steps.

\begin{claim}\label{claim_run_Delta_2M_from_v_to_v}
There is a run of type $(q_{\text{start}},1,q_{\text{finish}},0)$
on the pattern $\Delta_{2M}$
that first comes to some node $v_{i'}$, with $i' \leqslant n$,
in some state $\widehat{q}$,
having visited at most $n$ bottom elements $\Delta_2[*,\Delta_a]$ of $\Delta_{2M}$ by that time.
Next, the computation periodically repeats some sequence of proper steps,
with each proper step having pace strictly between $-2n$ and $2n$.
Every iteration of the period has pace at most $n$,
that is, it shifts the automaton by at most $n$ elements $\Delta_2[*,\Delta_a]$.
Furthermore, every iteration involves at most $2n$ consecutive elements $\Delta_2[*,\Delta_a]$.
The periodic part of the computation
leads the automaton to some node $v_{i''}$, with $i'' > 2M-n$,
in the same state $\widehat{q}$,
and after that the automaton finishes the run
visiting at most $n$ top elements $\Delta_2[*,\Delta_a]$.
\end{claim}
\begin{proof}
Consider the run through $\Delta_{2M}$
constructed in Claim~\ref{claim_run_Delta_2M_from_w_to_w},
which contains a periodic part.
Consider the first visit to the root port of some $\Delta_a$
in the periodically repeated sequence.
It takes place at some node $v_{i'}$ in some state $\widehat{q}$.
Then $i' \leqslant n$,
since, by Claim~\ref{claim_run_Delta_2M_from_w_to_w},
the prefix before the periodic part and the first iteration
together fit into the $n$ bottom elements $\Delta_2[*,\Delta_a]$.
The same visit exists in every iteration of the sequence,
and the number of the visited node is increased each time
by a certain fixed distance bounded by $n$.
The corresponding visit to the root port of $\Delta_a$ in the last iteration
takes place in some node $v_{i''}$ in the same state $\widehat{q}$.
Here $i'' > 2M-n$,
because, by Claim~\ref{claim_run_Delta_2M_from_w_to_w},
the last iteration and the suffix
fit into the $n$ top elements $\Delta_2[*,\Delta_a]$.

Next, the periodically repeated sequence is shifted
in order to begin with this visit to the root port of $\Delta_a$
(and accordingly to end with such a visit).
Since the old sequence is contained in $n$ consecutive elements $\Delta_2[*,\Delta_a]$,
the new sequence, like the old one, has pace at most $n$.
Furthermore, each iteration of the new sequence entirely fits
into two consecutive iterations of the original sequence,
and therefore is contained in $2n$ consecutive elements $\Delta_2[*,\Delta_a]$.
Then this new sequence can be split into segments
by the visits to the root ports of any $\Delta_a$.
These segments, however, are not necessarily proper steps in the sense of Definition~\ref{definition_step},
because they may contain multiple transfers through some elements $\Delta_2$,
possibly including returns back to the earlier elements and look-ahead into later elements,
as shown in Figure~\ref{f:move_i_p_q_gamma_runs}(left).
Nevertheless, each of these segments will be replaced with a proper step.

\begin{figure}[t]
	\centerline{\includegraphics[scale=0.9]{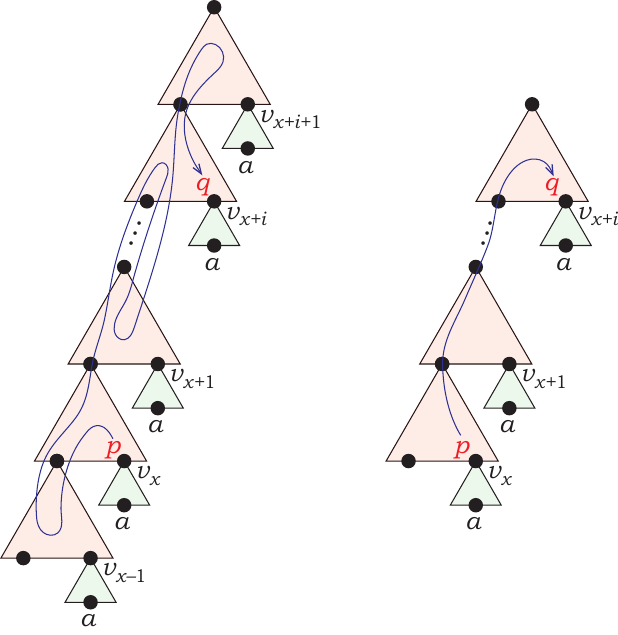}}
	\caption{(left) A computation from $(p, v_x)$ to $(q,v_{x+i})$
	that does not visit root ports of any $\Delta_a$ on the way;
	(right) its implementation by a proper step.}
	\label{f:move_i_p_q_gamma_runs}
\end{figure}

Consider one of these segments,
that is, a computation from the root port of some element $\Delta_a$
to the root port of another or the same $\Delta_a$,
without visiting root ports of any $\Delta_a$ on the way.
Let this segment begin in a node with some number $x$
and lead the automaton
from configuration $(p,v_x)$ to configuration $(q,v_{x+i})$,
where $i \in \mathbb{Z}$ is the difference between the node numbers,
$p$ is the state in the beginning of the segment,
and $q$ is the state in the end.
Because this segment is a part of the new periodically repeated sequence,
which fits into $2n$ consecutive elements $\Delta_2[*, \Delta_a]$,
the number $i$ must satisfy $-2n < i < 2n$.

In order to replace such a segment with a proper step,
the cases $i \neq 0$ and $i=0$ are considered separately.
Let $i \neq 0$.
Then the automaton never goes inside any elements $\Delta_a$ on this segment.
If all elements $\Delta_a$ are removed from the pattern $\Delta_{2M}$,
this results in a pattern $\Delta$ that consists of elements $\Delta_2$.
Its leaf ports are port $1$ at the bottom, and also all nodes $v_1, \ldots, v_{2M}$.
Then, by Lemma~\ref{lemma_gamma_simple_path}, the computation from $v_x$ to $v_{x+i}$
can be executed by a simple transfer, as in Figure~\ref{f:move_i_p_q_gamma_runs}(right),
which is a proper step.
Then the segment under consideration is replaced with this proper step.

Now consider the case $i = 0$.
In this case, by the first transition at the node $v_x$,
the automaton can either exit into the element $\Delta_2$,
or enter the element $\Delta_a$ in the subtree of $v_x$.
If the automaton enters $\Delta_a$,
then the whole segment is contained in this element $\Delta_a$,
that is, $p \circlearrowleft_a q$.
If the automaton exits into $\Delta_2$,
then it cannot enter any elements $\Delta_a$,
and therefore walks over the pattern $\Delta$
obtained from $\Delta_{2M}$ by removing all elements $\Delta_a$.
Then, by attaching an element $\Delta_1$ to the port $v_x$ of $\Delta$,
the automaton's computation from $(p,v_x)$ to $(q,v_x)$
becomes an inner loop by Lemma~\ref{lemma_inner_loops},
that is, $p \to_\epsilon q$.
In this case also $p \circlearrowleft_a q$.
Therefore, there is a proper step of type $(0, p, q)$,
with which this segment is replaced.

No proper step may go beyond its starting and ending elements $\Delta_2[*, \Delta_a]$.
Hence, the resulting periodically repeated sequence
is still contained in $2n$ consecutive elements $\Delta_2[*,\Delta_a]$.
\end{proof}

The next question is how proper steps of different types
pass by the \emph{fault}, that is,
the element $\Delta_2[*,\Delta_2[\Delta_a,\Delta_a]]$,
which distinguishes the faulty pattern $\Delta_{2M}'$
from the correct pattern $\Delta_{2M}$.

\section{Shrinking and stretching proper steps to bypass the fault}\label{section_stretched_and_shrunk_moves}

In this section it will be shown that proper steps
that pay attention to the fault in the faulty pattern $\Delta'_{2M}$
can be shrunk, or sometimes stretched,
so that they bypass the fault.

The possibility of shrinking is established in the following form:
if the automaton can move forward or backward by $i$ elements $\Delta_2[*,\Delta_a]$
on the correct pattern $\Delta_{2M}$,
then it can either do the same on the faulty pattern,
or it can move by $i-1$ elements on the correct pattern.

\begin{claim}\label{claim_shrunk_moves}
Assume that a proper step of type $(i,p,q)$ exists
for $-2n < i < 2n$, $i \neq 0$ and $p,q \in Q$.
Then at least one of two conditions holds:
\begin{enumerate}\renewcommand{\theenumi}{\Roman{enumi}}
\item	\label{claim_shrunk_moves__step_i_on_Delta_prime}
	for every integer $x$,
	such that $1 \leqslant x \leqslant 2M$ and $1 \leqslant x+i \leqslant 2M$, 
	the automaton $A$ can move on the faulty pattern $\Delta_{2M}'$ 
	from configuration $(p,v_x)$ to configuration $(q,v_{x+i})$;
\item	\label{claim_shrunk_moves__step_i_minus_1}
	there is a proper step of type $(i-1,p,q)$ for $i>0$,
	and of type $(i+1,p,q)$ for $i<0$.
\end{enumerate}
\end{claim}

\begin{proof}
Due to the time symmetry of the automaton $A$,
it is sufficient to consider only the case of $i > 0$
(the case of $i<0$ will be inferred from the case of positive $i$ in the end of the proof).

The cases $i > 1$ and $i = 1$ are considered separately.

\begin{figure}[t]
	\centerline{\includegraphics[scale=0.9]{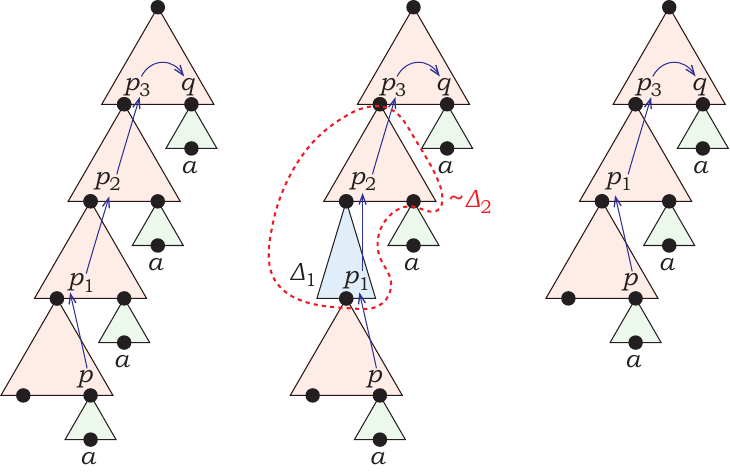}}
	\caption{%
	(left) proper step of type $(i,p,q)$, for $i = 3$;
	(middle) the second element $\Delta_2[*,\Delta_a]$ from the bottom replaced with $\Delta_1$;
	(right) the resulting proper step of type $(i-1,p,q)$.}
	\label{f:shrinking_move_case_i_greater_than_1}
\end{figure}

\begin{itemize}
\item
	In the case $i > 1$ it will be proved that
	a proper step of type $(i-1,p,q)$ exists.
	
	Consider the partition of any proper step of type $(i,p,q)$
	into transfers through elements $\Delta_2$.
	Let $p_1,\ldots,p_i$ be the intermediate states in this partition,
	as in Figure~\ref{f:shrinking_move_case_i_greater_than_1}(left).
	Then $p \nwarrow p_1 \nearrow p_2 \nearrow \ldots \nearrow p_i \curvearrowright q$
	by Claim~\ref{claim_sequence_for_step_partition}.
	By Lemma~\ref{lemma_delta1_up_and_down},
	$p_1 \nearrow p_2$ implies $p_1 \uparrow p_2$,
	that is, if the element $\Delta_2[*, \Delta_a]$ is replaced with the element $\Delta_1$,
	as in Figure~\ref{f:shrinking_move_case_i_greater_than_1}(middle),
	then the automaton can make a computation
	of the form $p \nwarrow p_1 \uparrow p_2 \nearrow \ldots \nearrow p_i \curvearrowright q$,
	without noticing the difference between the two patterns.
	
	Attaching an element $\Delta_1$ to any port of an element $\Delta_2$
	results in an element equivalent to $\Delta_2$.
	If $i>2$, then,
	by Lemma~\ref{lemma_swallowing_delta_1}(\ref{lemma_swallowing_delta_1__p_up_q_ne_r__p_ne_r}),
	$p_1 \uparrow p_2 \nearrow p_3$ implies $p_1 \nearrow p_3$.
	Then
	$p \nwarrow p_1 \nearrow p_3 \ldots \nearrow p_i \curvearrowright q$,
	and this is a proper step of type $(i-1,p,q)$,
	by Claim~\ref{claim_sequence_for_step_partition},
	see Figure~\ref{f:shrinking_move_case_i_greater_than_1}(right).
	
	If $i = 2$,
	then, by Lemma~\ref{lemma_swallowing_delta_1}(\ref{lemma_swallowing_delta_1__p_up_q_right_r__p_right_r}),
	$p_1 \uparrow p_2 \curvearrowright q$ implies $p_1 \curvearrowright q$,
	and there is a sequence $p \nwarrow p_1 \curvearrowright q$
	making a proper step of type $(i-1,p,q)$.
	
	Thus, Condition~\ref{claim_shrunk_moves__step_i_minus_1}
	holds in the case of $i > 1$.
\item
	In the case of $i = 1$,
	the goal of the proof is to show, for each number $x$,
	that either this $x$ satisfies Condition~\ref{claim_shrunk_moves__step_i_on_Delta_prime} 
	in the claim,
	or Condition~\ref{claim_shrunk_moves__step_i_minus_1} holds in general.
	Then, if Condition~\ref{claim_shrunk_moves__step_i_minus_1} is ever confirmed,
	the claim holds,
	and if it is never confirmed,
	Condition~\ref{claim_shrunk_moves__step_i_on_Delta_prime} holds for all $x$.
	
	Let $x$ be any integer with $1 \leqslant x \leqslant 2M-1$.
	It will be proved that either the automaton can move 
	on the faulty pattern $\Delta_{2M}'$ from configuration $(p,v_x)$ to configuration $(q,v_{x+1})$,
	or there exists a proper step of type $(0,p,q)$.
	
	A proper step of type $(1,p,q)$ splits into two transfers,
	that is, $p \nwarrow p_1 \curvearrowright q$ for some state $p_1$.
	Then, Lemma~\ref{lemma_delta1_up_and_down} asserts 
	that $p \uparrow p_1 \curvearrowright q$,
	from whence, by Lemma~\ref{lemma_swallowing_delta_1}(\ref{lemma_swallowing_delta_1__p_up_q_right_r__p_right_r}), $p \curvearrowright q$.
	A new pattern $\Delta$ of rank $2M+1$ is obtained out of the faulty pattern $\Delta'_{2M}$
	by removing all elements $\Delta_a$.
	Then the nodes $v_x$ and $v_{x+1}$ are consecutive ports
	of the pattern $\Delta$.
	Since $p \curvearrowright q$,
	by Lemma~\ref{lemma_transfers_through_Delta2_in_big_elements},
	either $(p,j,q,j+1)\in \gamma_\Delta$ for all $j \in \{1, \ldots, 2M\}$,
	or $p \to_\epsilon q$.
	In the former case, the automaton can move from configuration $(p,v_x)$
	to configuration $(q,v_{x+1})$ on the faulty pattern $\Delta'_{2M}$,
	and the number $x$ satisfies Condition~\ref{claim_shrunk_moves__step_i_on_Delta_prime}.
	And in the latter case there is a proper step of type $(0,p,q)$,
	and Condition~\ref{claim_shrunk_moves__step_i_minus_1} holds.
\end{itemize}

The case of positive $i$ has been proved.
Then the proof for the case of $i < 0$ can be obtained by the time symmetry of the automaton.
Applying the bijection $\tau$ that reverses the transitions,
as in Lemma~\ref{symmetries_lemma},
the states $p$ and $q$ are mapped to $\tau(p)$ and $\tau(q)$,
such that there is a proper step of type $(|i|,\tau(q),\tau(p))$.
Then, using the case of positive $i$,
either for all $x$ the automaton can move on the faulty pattern $\Delta_{2M}'$
from configuration $(\tau(q),v_{x+i})$ to configuration $(\tau(p),v_x)$,
or there is a proper step of type $(|i|-1,\tau(q),\tau(p))$.
In the former case, the time symmetry implies that, for all $x$,
there is a computation on the same faulty pattern from $(p,v_x)$ to $(q,v_{x+i})$,
whereas in the latter case, similarly, there is a proper step of type $(-|i|+1,p,q)$.
\end{proof}

The next claim is that sometimes proper steps can be not only shrunk,
but also stretched:
if the automaton can shift by $i$ elements upwards on the correct pattern $\Delta_{2M}$,
then on the faulty pattern $\Delta'_{2M}$
the automaton can either move upwards by the same distance without noticing the fault,
or it can move one element further,
or it can jump from anywhere to anywhere.

\begin{claim}\label{claim_stretched_moves}
Assume that there is a proper step of type $(i,p,q)$,
for some integer $i$ with $0 < i < 2n$, 
and for some two states $p,q \in Q$.
Let $x \geqslant 1$ be an integer bounded as $x+i \leqslant 2M$.
Then at least one of the following three conditions holds:
\begin{enumerate}\renewcommand{\theenumi}{\Roman{enumi}}
\item	\label{claim_stretched_moves__x_to_x_plus_i}
	the automaton $A$ can move on the faulty pattern $\Delta_{2M}'$ 
	from configuration $(p,v_x)$ to configuration $(q,v_{x+i})$;
\item	\label{claim_stretched_moves__x_to_x_plus_i_plus_1}
	the automaton can move on $\Delta_{2M}'$ 
	from configuration $(p,v_x)$ to configuration $(q,v_{x+i+1})$;
\item	\label{claim_stretched_moves__y_to_z}
	for all $y,z$, such that $1 \leqslant y < M$ and $M+1 < z \leqslant 2M$,
	the automaton can move
	on $\Delta_{2M}'$ from configuration $(p, v_y)$ to configuration $(q,v_z)$.
\end{enumerate}
\end{claim}

\begin{proof}
Let $p \nwarrow p_1 \nearrow p_2 \nearrow \ldots \nearrow p_i \curvearrowright q$,
for some intermediate states $p_1, \ldots, p_i \in Q$
in the partition of some proper step of type $(i,p,q)$
into transfers through elements $\Delta_2$,
as in Figure~\ref{f:stretching_move_case_x_less_M_x_plus_i_M}(left).
Since the faulty pattern $\Delta'_{2M}$ differs from the correct pattern $\Delta_{2M}$
only in a few elements constituting the fault,
a proper step of type $(i,p,q)$ can be repeated as it is in any positions far from the fault,
that is, if $x+i < M$ or $x>M+1$.
In this case Condition~\ref{claim_stretched_moves__x_to_x_plus_i} holds.

If the path from the node $v_x$ to the node $v_{x+i}$ comes near the fault,
that is, if $x \leqslant M \leqslant x+i$ or $x \leqslant M+1 \leqslant x+i$,
then there are five cases of relative position of the nodes $v_x$, $v_M$, $v_{M+1}$ and $v_{x+i}$.
These cases are considered separately.

\begin{figure}[t]
	\centerline{\includegraphics[scale=0.9]{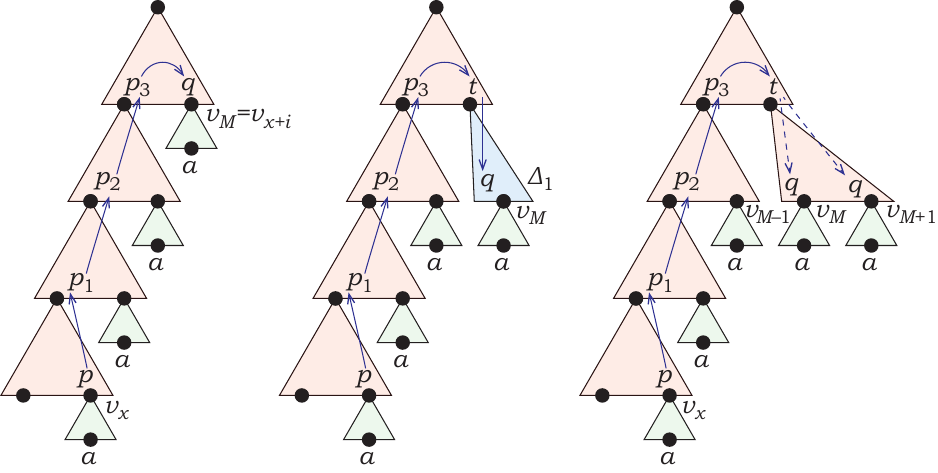}}
	\caption{Case $x+i=M$:
	(left) proper step of type $(i,p,q)$, for $i = 3$;
	(middle) $\Delta_1$ attached to the destination port;
	(right) on the faulty pattern $\Delta_{2M}'$, the automaton moving from $v_x$
	to $v_M$ (if $t \swarrow q$), or to $v_{M+1}$ ($t \searrow q$).}
	\label{f:stretching_move_case_x_less_M_x_plus_i_M}
\end{figure}

\begin{itemize}
\item
	Case $x < M$, $x+i = M$.
	
	By Lemma~\ref{lemma_inserting_delta_1}(\ref{lemma_inserting_delta_1_p_curvearrowright_q}),
	$p_i \curvearrowright q$ implies $p_i \curvearrowright t \downarrow q$,
	see Figure~\ref{f:stretching_move_case_x_less_M_x_plus_i_M}(middle).
	Next, by Lemma~\ref{lemma_delta1_up_and_down},
	there is at least one of the transfers $t \swarrow q$ and $t \searrow q$,
	as shown in Figure~\ref{f:stretching_move_case_x_less_M_x_plus_i_M}(right).
	If $t \swarrow q$,
	then the sequence of transfers
	$p \nwarrow p_1 \nearrow p_2 \nearrow \ldots \nearrow p_i \curvearrowright t \swarrow q$
	on the faulty pattern $\Delta_{2M}'$
	leads from configuration $(p,v_x)$ to configuration $(q,v_{M})$,
	and Condition~\ref{claim_stretched_moves__x_to_x_plus_i} holds.
	And if $t \searrow q$,
	then, similarly,
	$p \nwarrow p_1 \nearrow p_2 \nearrow \ldots \nearrow p_i \curvearrowright t \searrow q$,
	and the automaton moves on $\Delta_{2M}'$ from $(p,v_x)$ to $(q,v_{M+1})$;
	this is Condition~\ref{claim_stretched_moves__x_to_x_plus_i_plus_1}.

\begin{figure}[t]
	\centerline{\includegraphics[scale=0.9]{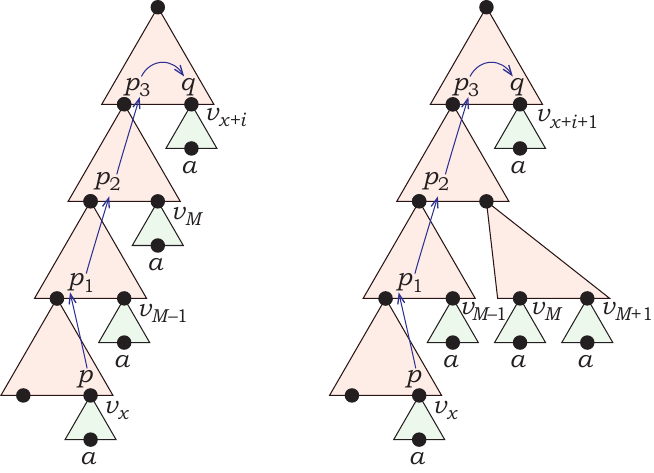}}
	\caption{Case $x < M$, $x+i \geqslant M+1$:
	(left) proper step of type $(i,p,q)$, with $i = 3$;
	(right) the same computation on the faulty pattern $\Delta_{2M}'$
	going from $v_x$ to $v_{x+i+1}$,
	skipping $\Delta_2[\Delta_a,\Delta_a]$ like an ordinary $\Delta_a$.}
	\label{f:stretching_move_case_x_less_M_x_plus_i_at_least_M_plus_1}
\end{figure}

\item
	Case $x < M$, $x+i \geqslant M+1$.
	
	Consider a proper step of type $(i,p,q)$,
	shown in Figure~\ref{f:stretching_move_case_x_less_M_x_plus_i_at_least_M_plus_1}(left).
	The same computation can be executed on the faulty pattern $\Delta_{2M}'$
	from the node $v_x$,
	as in Figure~\ref{f:stretching_move_case_x_less_M_x_plus_i_at_least_M_plus_1}(right),
	and the automaton bypasses the fault $\Delta_2[\Delta_a,\Delta_a]$
	as if it were $\Delta_a$.
	This computation leads the automaton
	from configuration $(p,v_x)$ to configuration $(q, v_{x+i+1})$.
	Thus, Condition~\ref{claim_stretched_moves__x_to_x_plus_i_plus_1} holds.

\begin{figure}[t]
	\centerline{\includegraphics[scale=0.9]{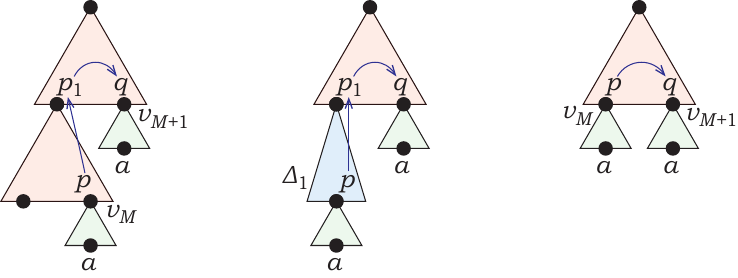}}
	\caption{Case $x=M$, $x+i=M+1$:
		(left) a proper step of type $(1,p,q)$;
		(middle) bottom element $\Delta_2$ replaced with $\Delta_1$;
		(right) a computation on the faulty pattern $\Delta_{2M}'$
			from $(p,v_M)$ to $(q,v_{M+1})$.}
	\label{f:stretching_move_case_x_is_M_x_plus_i_is_M_plus_1}
\end{figure}

\item
	Case $x = M$, $x+i = M+1$.
	
	In this case, by Lemma~\ref{lemma_delta1_up_and_down},
	$p \nwarrow p_1 \curvearrowright q$,
	implies $p \uparrow p_1 \curvearrowright q$,
	as shown in Figure~\ref{f:stretching_move_case_x_is_M_x_plus_i_is_M_plus_1}(left,middle).
	Next, by Lemma~\ref{lemma_swallowing_delta_1}(\ref{lemma_swallowing_delta_1__p_up_q_right_r__p_right_r}),
	$p \curvearrowright q$,
	and the automaton moves on the faulty pattern $\Delta_{2M}'$
	from configuration $(p,v_M)$ to configuration $(q, v_{M+1})$,
	as in Figure~\ref{f:stretching_move_case_x_is_M_x_plus_i_is_M_plus_1}(right).
	Hence, Condition~\ref{claim_stretched_moves__x_to_x_plus_i} is satisfied.

\begin{figure}[t]
	\centerline{\includegraphics[scale=0.9]{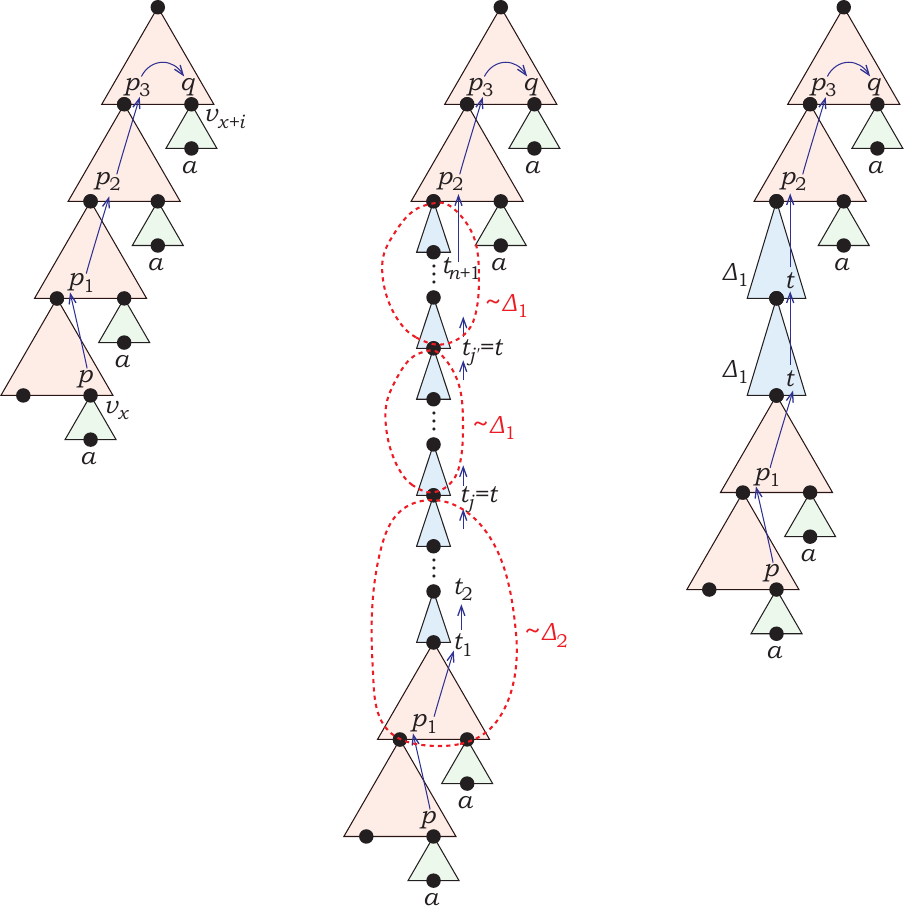}}
	\caption{Case $x = M$, $x+i > M+1$:
	(left) a proper step of type $(i,p,q)$, for $i=3$;
	(middle) inserting a chain of $n+1$ elements $\Delta_1$;
	(right) elements $\Delta_1$ merged with the neighbouring elements.}
	\label{f:stretching_move_case_x_is_M_x_plus_i_more_M_plus_1}
\end{figure}

\item
	Case $x = M$, $x+i > M+1$.
	
	A proper step of type $(i,p,q)$ is a sequence
	$p \nwarrow p_1 \nearrow p_2 \nearrow \ldots \nearrow p_i \curvearrowright q$,
	illustrated in Figure~\ref{f:stretching_move_case_x_is_M_x_plus_i_more_M_plus_1}(left).
	By Lemma~\ref{lemma_inserting_delta_1}(\ref{lemma_inserting_delta_1_p_nearrow_q}),
	one can insert a chain of $n+1$ elements $\Delta_1$,
	as shown in Figure~\ref{f:stretching_move_case_x_is_M_x_plus_i_more_M_plus_1}(middle),
	resulting in
	$p \nwarrow p_1 \nearrow t_1 \uparrow t_2 \uparrow \ldots \uparrow t_{n+1} \uparrow p_2 \nearrow \ldots \nearrow p_i \curvearrowright q$,
	for some intermediate states $t_1, \ldots, t_{n+1} \in Q$.
	Since $|Q| = n$, among the states $t_1, \ldots, t_{n+1}$
	some two states coincide: $t_j=t_{j'}$;
	this repeated state is denoted by $t$.
	Then, by Lemma~\ref{lemma_swallowing_delta_1}(\ref{lemma_swallowing_delta_1__p_up_q_up_r__p_up_r},\ref{lemma_swallowing_delta_1__p_ne_q_up_r__p_ne_r}),
	almost all $\Delta_1$ inserted
	can be merged into the neighbouring elements,
	so that one $\Delta_1$ remains above the state $t_{j'}$,
	one between $t_j$ and $t_{j'}$,
	and the rest of them are merged with $\Delta_2$ below.
	This results in the sequence
	$p \nwarrow p_1 \nearrow t \uparrow t \uparrow p_2 \nearrow \ldots \nearrow p_i \curvearrowright q$,
	see Figure~\ref{f:stretching_move_case_x_is_M_x_plus_i_more_M_plus_1}(right).
	
	By Lemma~\ref{lemma_delta1_up_and_down},
	either $t \nearrow t$, or $t \nwarrow t$.
	These two cases are considered separately.
	
\begin{figure}[t]
	\centerline{\includegraphics[scale=0.9]{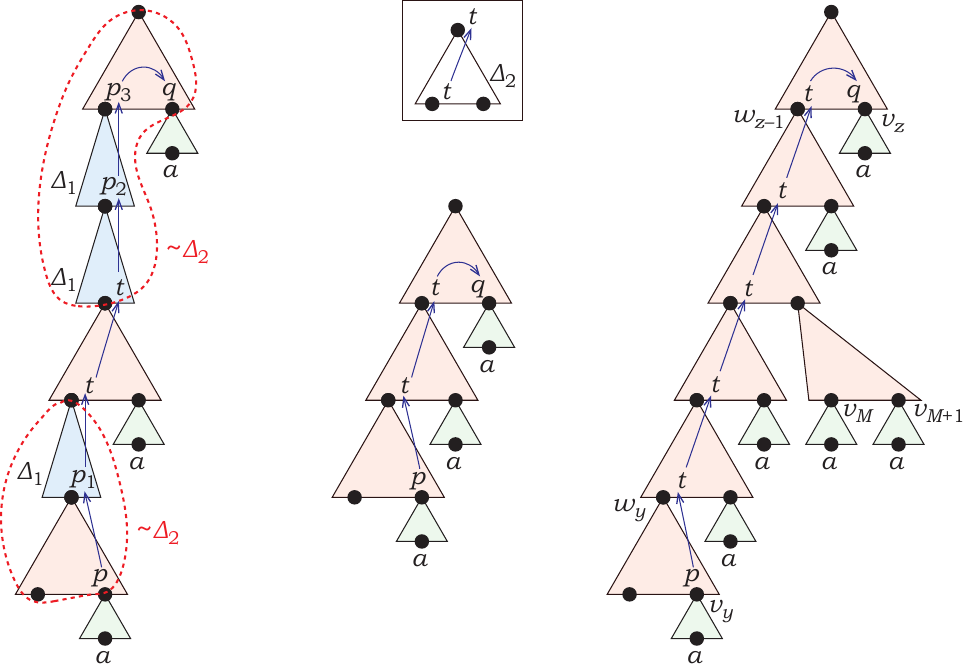}}
	\caption{Continuation of Figure~\ref{f:stretching_move_case_x_is_M_x_plus_i_more_M_plus_1}
	for $t \nearrow t$:
	(left) some elements $\Delta_2$ replaced with $\Delta_1$ and vice versa;
	(middle) all elements $\Delta_1$ merged with the neighbouring elements;
	(right) repeating $t \nearrow t$ to move from $v_y$ to $v_z$ 
	on the faulty pattern $\Delta_{2M}'$.}
	\label{f:stretching_move_case_x_is_M_x_plus_i_more_M_plus_1_first_part}
\end{figure}

	First, let $t \nearrow t$.
	This allows the following modifications to the computation
	in Figure~\ref{f:stretching_move_case_x_is_M_x_plus_i_more_M_plus_1}(right):
	first, the lower of the two elements $\Delta_1$
	(the one with $t \uparrow t$) is replaced with $\Delta_2$ (with $t \nearrow t$);
	secondly, all original $\Delta_2$ except for the top and the bottom ones,
	are replaced with $\Delta_1$ using Lemma~\ref{lemma_delta1_up_and_down}.
	This yields the sequence
	$p \nwarrow p_1 \uparrow t \nearrow t \uparrow p_2 \uparrow \ldots \uparrow p_i \curvearrowright q$,
	illustrated in Figure~\ref{f:stretching_move_case_x_is_M_x_plus_i_more_M_plus_1_first_part}(left).
	Next, Lemma~\ref{lemma_swallowing_delta_1}(\ref{lemma_swallowing_delta_1__p_nw_q_up_r__p_nw_r},\ref{lemma_swallowing_delta_1__p_up_q_right_r__p_right_r})
	allows the elements $\Delta_1$ to be merged with any neighbouring elements,
	and the sequence $p \nwarrow t \nearrow t \curvearrowright q$ can be thus obtained,
	see Figure~\ref{f:stretching_move_case_x_is_M_x_plus_i_more_M_plus_1_first_part}(middle).
	Then, for all $y,z$, with $1\leqslant y < M$ and $M+1 < z \leqslant 2M$,
	the automaton can move on the faulty pattern $\Delta_{2M}'$ 
	from configuration $(p, v_y)$ to configuration $(q,v_z)$,
	as in Figure~\ref{f:stretching_move_case_x_is_M_x_plus_i_more_M_plus_1_first_part}(right).
	Indeed, one can first make the transfer $p \nwarrow t$,
	to come to the node $w_y$ (the root of $\Delta_2$),
	then repeat transfers $t \nearrow t$, 
	skipping all $\Delta_a$, as well as the fault $\Delta_2[\Delta_a,\Delta_a]$,
	until the automaton comes to the node $w_{z-1}$;
	and finally one can use $t \curvearrowright q$
	to come to the destination configuration $(q,v_z)$.
	Thus, Condition~\ref{claim_stretched_moves__y_to_z} holds in this case.

\begin{figure}[t]
	\centerline{\includegraphics[scale=0.9]{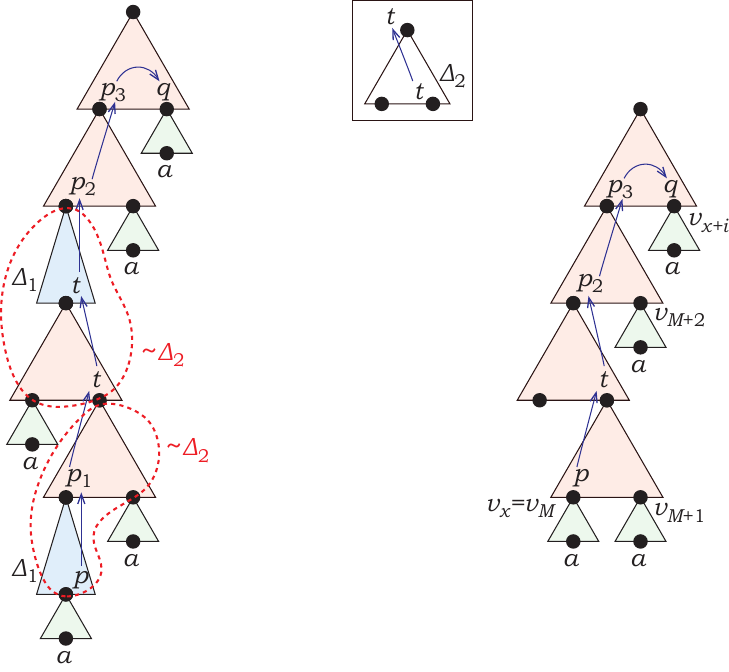}}
	\caption{Continuation of Figure~\ref{f:stretching_move_case_x_is_M_x_plus_i_more_M_plus_1}
	for $t \nwarrow t$:
	(left) one $\Delta_2$ replaced with $\Delta_1$, and one $\Delta_1$ with $\Delta_2$;
	(right) the automaton moves from $v_M$ to $v_{M+i}$ on the faulty pattern $\Delta_{2M}'$.}
	\label{f:stretching_move_case_x_is_M_x_plus_i_more_M_plus_1_second_part}
\end{figure}

	Now let $t \nwarrow t$.
	Then, in the pattern
	in Figure~\ref{f:stretching_move_case_x_is_M_x_plus_i_more_M_plus_1}(right),
	the lower of the two elements $\Delta_1$ (with $t \uparrow t$)
	is replaced with $\Delta_2$ (with $t \nwarrow t$),
	and also the bottom element $\Delta_2$ is replaced with $\Delta_1$,
	by Lemma~\ref{lemma_delta1_up_and_down}.
	Then $p \uparrow p_1 \nearrow t \nwarrow t \uparrow p_2 \nearrow \ldots \nearrow p_i \curvearrowright q$,
	see Figure~\ref{f:stretching_move_case_x_is_M_x_plus_i_more_M_plus_1_second_part}(left).
	Next, by Lemma~\ref{lemma_swallowing_delta_1}(\ref{lemma_swallowing_delta_1__p_up_q_ne_r__p_ne_r},\ref{lemma_swallowing_delta_1__p_nw_q_up_r__p_nw_r}),
	both elements $\Delta_1$ are merged into the neighbouring elements $\Delta_2$,
	resulting in the sequence
	$p \nearrow t \nwarrow p_2 \nearrow \ldots \nearrow p_i \curvearrowright q$.
	And this sequence leads the automaton $A$ on the faulty pattern $\Delta_{2M}'$
	from configuration $(p,v_M)$ to configuration $(q,v_{M+i})$, 
	as in Figure~\ref{f:stretching_move_case_x_is_M_x_plus_i_more_M_plus_1_second_part}(right).
	This is Condition~\ref{claim_stretched_moves__x_to_x_plus_i}.
	
\begin{figure}[t]
	\centerline{\includegraphics[scale=0.9]{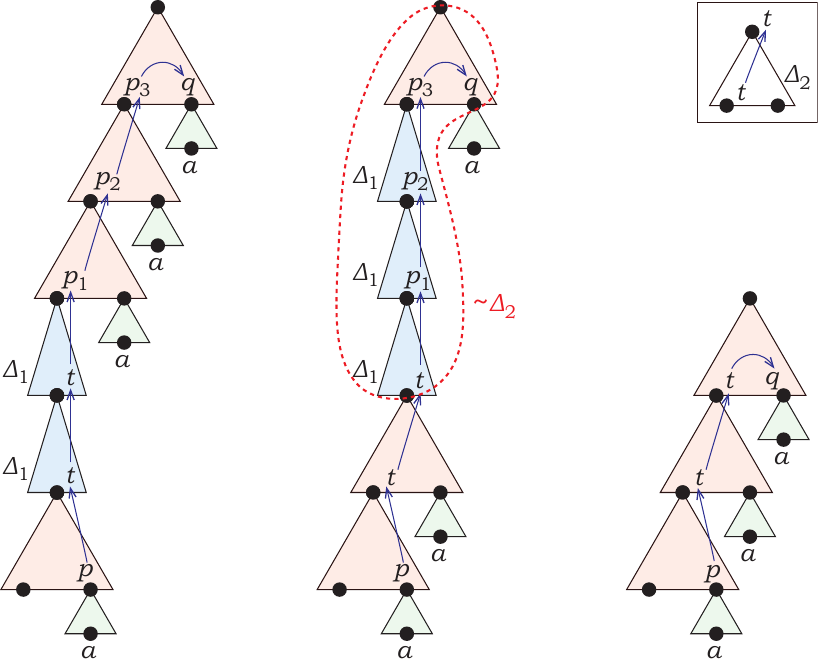}}
	\caption{Case $x = M+1$, $x+i > M+1$:
	(left) a computation obtained from a proper step of type $(i,p,q)$, for $i=3$;
	(middle) replacing some elements $\Delta_2$ with $\Delta_1$ and vice versa, for $t \nearrow t$;
	(right) all elements $\Delta_1$ merged with the neighbouring $\Delta_2$,
	obtaining a computation in which $t \nearrow t$ can be repeated.}
	\label{f:stretching_move_case_x_is_M_plus_1_first_case}
\end{figure}

\item
	Case $x = M+1$, $x+i > M+1$.
	
	This case is very similar to the previous one.
	For a proper step $(i,p,q)$, which is of the form,
	$p \nwarrow p_1 \nearrow p_2 \nearrow \ldots \nearrow p_i \curvearrowright q$,
	as in Figure~\ref{f:stretching_move_case_x_is_M_x_plus_i_more_M_plus_1}(left),
	a chain of $n+1$ elements $\Delta_1$ is similarly inserted,
	but the insertion point is now below $p_1$
	(cf.\ below $p_2$ in the previous case).
	This is done by Lemma~\ref{lemma_inserting_delta_1}(\ref{lemma_inserting_delta_1_p_nwarrow_q}),
	and the resulting sequence is 
	$p \nwarrow t_1 \uparrow t_2 \uparrow \ldots \uparrow t_{n+1} \uparrow p_1 \nearrow p_2 \nearrow \ldots \nearrow p_i \curvearrowright q$,
	where $t_1, \ldots, t_{n+1} \in Q$ are the intermediate states,
	with $t_j = t_{j'} = t$, for some $1 \leqslant j < j' \leqslant n+1$.
	And next, by Lemma~\ref{lemma_swallowing_delta_1}(\ref{lemma_swallowing_delta_1__p_up_q_up_r__p_up_r},\ref{lemma_swallowing_delta_1__p_nw_q_up_r__p_nw_r}),
	one obtains the sequence 
	$p \nwarrow t \uparrow t \uparrow p_1 \nearrow p_2 \nearrow \ldots \nearrow p_i \curvearrowright q$,
	which is shown in Figure~\ref{f:stretching_move_case_x_is_M_plus_1_first_case}(left).
	As before, either $t \nearrow t$, or $t \nwarrow t$.
	
	If $t \nearrow t$,
	then, as in the previous case,
	the computation in Figure~\ref{f:stretching_move_case_x_is_M_plus_1_first_case}(left)
	is modified as follows:
	the bottom $\Delta_1$ is replaced with $\Delta_2$ (with $t \nearrow t$),
	and all original $\Delta_2$ except the top and the bottom ones
	are replaced with $\Delta_1$ using Lemma~\ref{lemma_delta1_up_and_down}.
	This results in the sequence
	$p \nwarrow t \nearrow t \uparrow p_1 \uparrow p_2 \uparrow \ldots \uparrow p_i \curvearrowright q$,
	in Figure~\ref{f:stretching_move_case_x_is_M_plus_1_first_case}(middle).
	Next, by Lemma~\ref{lemma_swallowing_delta_1}(\ref{lemma_swallowing_delta_1__p_up_q_right_r__p_right_r}),
	all elements $\Delta_1$ are merged with the top $\Delta_2$,
	giving the sequence $p \nwarrow t \nearrow t \curvearrowright q$,
	see Figure~\ref{f:stretching_move_case_x_is_M_plus_1_first_case}(right).
	Finally, as in the previous case,
	this sequence implies Condition~\ref{claim_stretched_moves__y_to_z}.

\begin{figure}[t]
	\centerline{\includegraphics[scale=0.9]{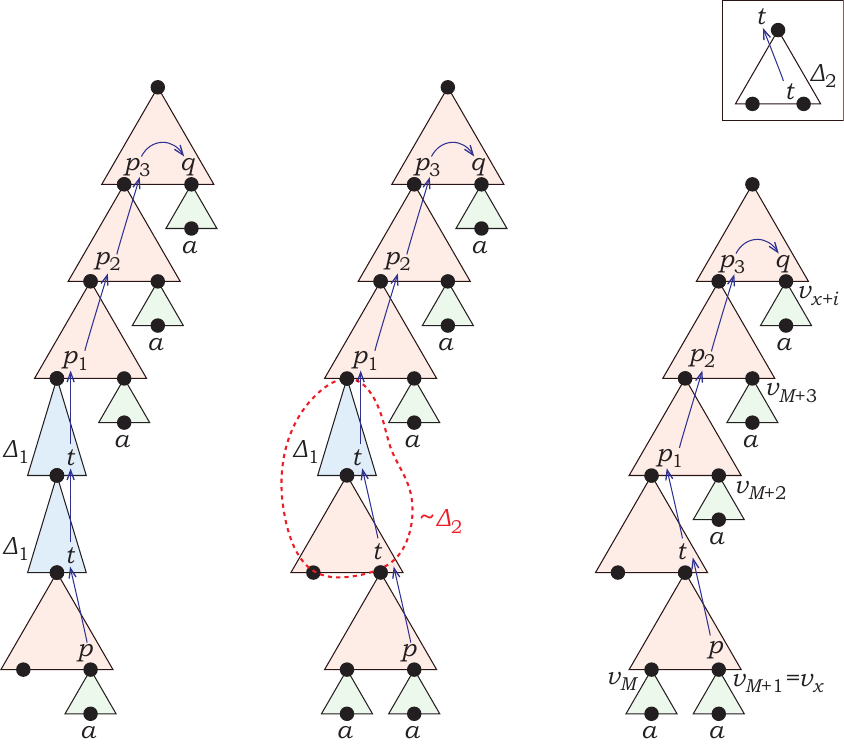}}
	\caption{Case $x = M+1$, $x+i > M+1$:
	(left) same computation as in Figure~\ref{f:stretching_move_case_x_is_M_plus_1_first_case},
	obtained from a proper step of type $(i,p,q)$, for $i=3$;
	(middle) one element $\Delta_1$ replaced with $\Delta_2$, for $t \nwarrow t$;
	(right) the remaining $\Delta_1$ combined with $\Delta_2$,
	and the resulting computation from $(p,v_{M+1})$ to $(q,v_{M+1+i})$ on the faulty pattern $\Delta_{2M}'$.}
	\label{f:stretching_move_case_x_is_M_plus_1_second_case}
\end{figure}

	Now let $t \nwarrow t$.
	In this case, the lower element $\Delta_1$ can be replaced with $\Delta_2$ (with $t \nwarrow t$),
	forming the sequence
	$p \nwarrow t \nwarrow t \uparrow p_1 \nearrow p_2 \nearrow \ldots \nearrow p_i \curvearrowright q$,
	shown in Figure~\ref{f:stretching_move_case_x_is_M_plus_1_second_case}(middle).
	Next, Lemma~\ref{lemma_swallowing_delta_1}(\ref{lemma_swallowing_delta_1__p_nw_q_up_r__p_nw_r})
	allows the remaining element $\Delta_1$ to be merged into the $\Delta_2$ below,
	resulting in the sequence
	$p \nwarrow t \nwarrow p_1 \nearrow p_2 \nearrow \ldots \nearrow p_i \curvearrowright q$.
	Using this sequence, the automaton $A$ operating on the faulty pattern $\Delta_{2M}'$
	moves from configuration $(p,v_{M+1})$ to configuration $(q,v_{M+1+i})$,
	as in Figure~\ref{f:stretching_move_case_x_is_M_plus_1_second_case}(right).
	Condition~\ref{claim_stretched_moves__x_to_x_plus_i} is met in this case.
\end{itemize}
\end{proof}

\section{How to move through the faulty pattern $\Delta_{2M}'$ without noticing the fault}\label{section_passing_through_Delta_prime_without_noticing_error}

In this section,
the proof of the Main Lemma will be completed,
along with the whole proof of Theorem~\ref{ntwa_rec_L_but_not_compl_of_L}
stating that the nondeterministic tree-walking automata are not closed under complementation.
It remains to prove that
$(q_{\text{start}},1,q_{\text{finish}},0) \in \delta_{\Delta_{2M}'}$,
using Claims~\ref{claim_run_Delta_2M_from_v_to_v},~\ref{claim_shrunk_moves}
and~\ref{claim_stretched_moves}.

The desired computation
from state $q_{\text{start}}$ in the leaf port of the faulty pattern $\Delta_{2M}'$
to state $q_{\text{finish}}$ in its root port
is constructed as follows.
Consider the computation on the correct pattern $\Delta_{2M}$
from Claim~\ref{claim_run_Delta_2M_from_v_to_v}:
first, the automaton moves from $q_{\text{start}}$ in the leaf port
to the node $v_{i'}$, for some $i' \leqslant n$, in some state $\widehat{q}$,
having visited only the bottom $n$ elements $\Delta_2[*,\Delta_a]$.
Next, the automaton $A$ repeats periodically
some sequence of proper steps, each with pace strictly between $-2n$ and $2n$.
Each iteration of the repeated sequence
is contained in some $2n$ consecutive elements $\Delta_2[*,\Delta_a]$.
Eventually the automaton comes to some node $v_{i''}$, for some $i'' > 2M-n$, in the same state $\widehat{q}$.
And then it finishes its computation
in the root port in the state $q_{\text{finish}}$,
while visiting only the top $n$ elements $\Delta_2[*,\Delta_a]$.
The computation of the automaton $A$ on the correct pattern $\Delta_{2M}$ described above
is fixed for the rest of this section.
And now the goal is to modify this computation
to reproduce it on the faulty pattern $\Delta_{2M}'$.

Since $M > n$, the faulty pattern $\Delta_{2M}'$
begins and ends with $n$ elements $\Delta_2[*,\Delta_a]$,
just like the correct pattern $\Delta_{2M}$.
Therefore, the automaton $A$, working on $\Delta_{2M}'$,
can repeat the first and the last parts of the original computation:
it can move up to configuration $(\widehat{q},v_{i'})$,
and it can finish the computation from configuration $(\widehat{q},v_{i''})$.
It remains to prove that the automaton $A$
can also move from configuration $(\widehat{q},v_{i'})$
to configuration $(\widehat{q},v_{i''})$.
on the faulty pattern $\Delta_{2M}'$.

The next claim considers the simple case
when one of the proper steps in the periodically repeated sequence
can be stretched to almost the entire faulty pattern $\Delta'_{2M}$.

\begin{claim} \label{claim_eliminating_yz_case}
Assume that one of the proper steps in the periodically repeated sequence
has type $(i,p,q)$, with $0 < i < 2n$ and $p,q \in Q$,
and satisfies Condition~\ref{claim_stretched_moves__y_to_z}
from Claim~\ref{claim_stretched_moves};
that is, for all $y$ and $z$,
with $1 \leqslant y < M$ and $M+1 < z \leqslant 2M$,
the automaton $A$ can move on the faulty pattern $\Delta_{2M}'$
from configuration $(p, v_y)$ to configuration $(q,v_z)$.
Then $A$ can move on $\Delta_{2M}'$
from configuration $(\widehat{q},v_{i'})$
to configuration $(\widehat{q},v_{i''})$.
\end{claim}
\begin{proof}
Consider the part of the computation of $A$ on $\Delta_{2M}$
starting from $(\widehat{q},v_{i'})$.
A sequence of proper steps is repeated periodically,
and a proper step of type $(i,p,q)$ occurs in every iteration.
Let $(p,v_y)$ be the configuration in which this proper step begins in the first iteration.
Then $y \leqslant i'+2n \leqslant 3n < M$.
Similarly, let $(q,v_z)$ be the configuration in which this proper step ends in the last iteration.
Then $z \geqslant i''-2n > 2M-3n > M+1$.

Note that, since each iteration moves the automaton upwards
and is contained in $2n$ consecutive elements $\Delta_2[*,\Delta_a]$,
if the automaton comes to some node $v_j$ during its periodic computation,
then it shall not return to any nodes from $v_1$ to $v_{j-2n}$,
and has not yet been to any nodes from $v_{j+2n}$ to $v_{2M}$.

Then there is the following computation on $\Delta_{2M}'$,
proceeding from configuration $(\widehat{q},v_{i'})$
to configuration $(\widehat{q},v_{i''})$.
First the automaton comes to the configuration $(p,v_y)$ as on the correct pattern $\Delta_{2M}$;
this can be done, because the path to configuration $(p,v_y)$
visits at most $y+2n \leqslant 5n < M$ 
bottom elements $\Delta_2[*,\Delta_a]$,
and hence cannot distinguish $\Delta_{2M}'$ from $\Delta_{2M}$.
Next, the automaton $A$ moves from configuration $(p,v_y)$ to configuration $(q, v_z)$:
this can be done by the assumption.
Finally, the automaton comes to configuration $(\widehat{q},v_{i''})$,
repeating the computation on $\Delta_{2M}$;
this last part of the computation visits only elements $\Delta_2[*,\Delta_a]$
with numbers at least $z-2n > 2M-5n > M+1$,
and hence can be executed on the faulty pattern $\Delta_{2M}'$.
\end{proof}

Thus, if a proper step
satisfying Condition~\ref{claim_stretched_moves__y_to_z} from Claim~\ref{claim_stretched_moves}
is used in the periodically repeated part of the computation,
then the automaton $A$ can move on the faulty pattern $\Delta_{2M}'$
from configuration $(\widehat{q},v_{i'})$ to configuration $(\widehat{q},v_{i''})$,
and the proof of the Main Lemma in this case is completed.
In the rest of the proof,
it is assumed that there are no such proper steps in the periodically repeated sequence.

The next claim is that if the automaton $A$
works on the faulty pattern $\Delta_{2M}'$,
then it may bypass the fault and get to some node $v_j$ after the fault
in the state $\widehat{q}$.
It will come not necessarily to one of the nodes
to which the original computation on the correct pattern $\Delta_{2M}$
arrives in the state $\widehat{q}$;
what is important is that it comes in this state
to \emph{some} node far beyond the fault.

\begin{claim}\label{claim_passing_by_v_M_v_M_plus_1}
The automaton $A$ operating on the faulty pattern $\Delta_{2M}'$
may move from configuration $(\widehat{q},v_{i'})$
to some configuration $(\widehat{q}, v_j)$,
where $j$ is a number with $M+2n+1 < j < M+8n$.
\end{claim}
\begin{proof}
The idea of the proof is to take each proper step
used in the periodic part of the computation on the correct pattern,
and to reproduce it in the computation on the faulty pattern.
When this is impossible, that proper step will be modified:
a proper step downward will be shrunk using Claim~\ref{claim_shrunk_moves},
and an upward proper step will be stretched by Claim~\ref{claim_stretched_moves}.
This way, the sequence of modified proper steps on the faulty pattern $\Delta'_{2M}$
will lead the automaton upward faster than
the original sequence of proper steps on the correct pattern. 
Finally, once the modified sequence passes by the fault,
the desired configuration $(\widehat{q}, v_j)$
can be obtained by just finishing the current iteration.

Let $\ell$ be the number of proper steps in the periodic part of the computation on $\Delta_{2M}$, 
which is the part of the computation between configurations
$(\widehat{q},v_{i'})$ and $(\widehat{q},v_{i''})$.
Consider the sequence of configurations 
$(p_0,v_{j_0})$, $(p_1,v_{j_1})$, \ldots, $(p_\ell,v_{j_\ell})$,
where $(p_0,v_{j_0}) = (\widehat{q}, v_{i'})$,
and all configurations except the first one
are entered on $\Delta_{2M}$ after making proper steps in the periodic part. 
Then, $(p_\ell,v_{j_\ell}) = (\widehat{q}, v_{i''})$.

The proof is by constructing a corresponding sequence of configurations
$(p_0,v_{j'_0})$, $(p_1,v_{j'_1})$, \ldots, $(p_m, v_{j'_m})$
on the faulty pattern $\Delta'_{2M}$, with $m \leqslant \ell$,
which starts in the same configuration $(p_0,v_{j'_0}) = (p_0,v_{j_0}) = (\widehat{q}, v_{i'})$,
and then passes through the same states,
while accumulating some non-negative deviation from the original sequence.
The new sequence ends as soon as it bypasses the fault
by a certain margin: $m$ is the least number with $j'_m > M+6n$.
The sequence should satisfy the following conditions:
\begin{itemize}
\item
	for each $k$, with $0 \leqslant k \leqslant m$,
	the automaton $A$ on $\Delta'_{2M}$
	can move from configuration $(\widehat{q}, v_{i'})$
	to configuration $(p_k,v_{j'_k})$;
\item
	each node $v_{j'_k}$ either equals the corresponding node $v_{j_k}$
	in the computation on $\Delta_{2M}$,
	or is closer to the root,
	that is, $j'_k \geqslant j_k$;
\item
	each node in the sequence is not too far from the previous one:
	$|j'_k - j'_{k-1}| \leqslant 2n$.
\end{itemize}
Such configurations on $\Delta'_{2M}$ are constructed inductively.
The base case is $k=0$, for which $(p_0,v_{j'_0}) = (p_0,v_{j_0})$:
no deviation has been accumulated so far.

For the induction step,
let $k > 0$, and assume that a sequence of configurations
$(p_0,v_{j'_0})$, \ldots, $(p_{k-1},v_{j'_{k-1}})$ satisfying the above properties
has been constructed,
and all these configurations are in nodes with numbers at most $M+6n$.
The goal is to construct the next configuration $(p_k,v_{j'_k})$.

On the correct pattern, the automaton moves
from configuration $(p_{k-1},v_{j_{k-1}})$ to configuration $(p_k,v_{j_k})$,
making a proper step of some type $(i_k,p_{k-1},p_k)$, with $-2n < i_k < 2n$.

On the faulty pattern $\Delta_{2M}'$,
if the automaton can move from configuration $(p_{k-1},v_{j_{k-1}'})$
to configuration $(p_k,v_{j_{k-1}'+i_k})$,
then one can take $j_k' = j_{k-1}'+i_k$,
and $j_k = j_{k-1}+i_k \leqslant j'_{k-1}+i_k = j_k'$,
and $|j'_k-j'_{k-1}| = |i_k| < 2n$,
and the induction step is proved in this case.

Now assume that the automaton cannot move on the faulty pattern $\Delta_{2M}'$ 
from $(p_{k-1},v_{j_{k-1}'})$ to $(p_k,v_{j_{k-1}'+i_k})$.
Then, $i_k \neq 0$,
because a proper step of pace $0$ is a transfer through $\Delta_a$,
and it can be made everywhere on the faulty pattern $\Delta_{2M}'$. 

Consider the cases of negative and positive $i_k$.
The first case is $i_k < 0$, that is, of a proper step backwards.
Then Claim~\ref{claim_shrunk_moves} is applicable,
and Condition~\ref{claim_shrunk_moves__step_i_on_Delta_prime} therein does not hold
by the assumption that on the faulty pattern $\Delta_{2M}'$
the automaton cannot move from $(p_{k-1},v_{j_{k-1}'})$ to $(p_k,v_{j_{k-1}'+i_k})$.
Therefore, Condition~\ref{claim_shrunk_moves__step_i_minus_1} of Claim~\ref{claim_shrunk_moves} holds,
that is, there is a proper step of type $(i_k+1,p_{k-1},p_k)$, which is shorter by one.
If for this new proper step the automaton again cannot move on the faulty pattern $\Delta_{2M}'$
from $(p_{k-1},v_{j_{k-1}'})$ to $(p_k,v_{j_{k-1}'+i_k+1})$,
then Claim~\ref{claim_shrunk_moves} is applied again,
Condition~\ref{claim_shrunk_moves__step_i_on_Delta_prime} fails again,
and a proper step of type $(i_k+2,p_{k-1},p_k)$ exists.
Thus, using Claim~\ref{claim_shrunk_moves},
one can reduce the absolute value of the pace
until a shorter proper step from this node $v_{j_{k-1}'}$
can be adapted for the faulty pattern $\Delta_{2M}'$.
And this will eventually happen
because a step of pace $0$ can be executed everywhere on the faulty pattern $\Delta_{2M}'$.

The conclusion is that
on the faulty pattern $\Delta_{2M}'$
the automaton can move from $(p_{k-1},v_{j_{k-1}'})$
to $(p_k,v_{j_{k-1}'+i'_k})$, for some $i'_k \leqslant 0$ with $|i'_k|<|i_k|$.
Let $j'_k=j'_{k-1}+i'_k$. Then, $j_k = j_{k-1} + i_k \leqslant j'_{k-1}+i_k < j'_{k-1}+i'_k = j'_k$,
and also $|j'_k-j'_{k-1}| = |i'_k| < |i_k| <2n$.

The second case is $i_k > 0$.
By assumption, a proper step of type $(i_k,p_{k-1},p_k)$
does not satisfy Condition~\ref{claim_stretched_moves__y_to_z} from Claim~\ref{claim_stretched_moves}.
Condition~\ref{claim_stretched_moves__x_to_x_plus_i} from Claim~\ref{claim_stretched_moves}
does not hold either, since it is assumed
that the automaton cannot move
from configuration $(p_{k-1},v_{j_{k-1}'})$ to configuration $(p_k,v_{j_{k-1}'+i_k})$
on the faulty pattern $\Delta_{2M}'$.
Thus, Condition~\ref{claim_stretched_moves__x_to_x_plus_i_plus_1} in Claim~\ref{claim_stretched_moves} holds,
and the automaton can move
from $(p_{k-1},v_{j_{k-1}'})$ to $(p_k,v_{j_{k-1}'+i_k+1})$,
that is, by one element farther.
Therefore, one can take $j_k' = j_{k-1}'+i_k+1$,
and then, first, $j_k = j_{k-1}+i_k \leqslant j_{k-1}'+i_k < j_{k-1}'+i_k+1 = j_k'$,
and secondly, $|j'_k-j_{k-1}'| = i_k+1 \leqslant 2n-1+1 \leqslant 2n$.

Thus, each node in the new sequence of configurations on the faulty pattern
is either closer to the root
than the corresponding node of the original sequence on the correct pattern,
or is the same node.
Hence, sooner or later the new sequence will advance beyond position $M+6n$.
This completes the construction of the new sequence.

Thus, a sequence of configurations
$(p_0,v_{j'_0})$, \ldots, $(p_m,v_{j'_m})$
on the faulty pattern $\Delta'_{2M}$ has been constructed:
all of them are reachable from the configuration $(\widehat{q},v_{i'})$,
the distance between any two of them does not exceed $2n$,
and all of them except the last one are not above position $M+6n$.
Then, for the next to last configuration $(p_{m-1},v_{j'_{m-1}})$,
the following inequality holds: $M+4n+1 \leqslant j'_{m-1} \leqslant M+6n$.
From position $j'_{m-1}$,
no iterations of the period may lead the automaton back to the fault,
because each iteration is contained in $2n$ consecutive elements $\Delta_2[*,\Delta_a]$.
Therefore, the current iteration can be completed from state $p_{m-1}$
to state $\widehat{q}$, so that the automaton gets to the desired configuration $(\widehat{q},v_j)$,
with $|j-j'_{m-1}| < 2n$, and hence $M+2n+1 < j < M+8n$.
\end{proof}

Now everything is prepared for the final step of the proof of the Main Lemma.
It has been proved that on the faulty pattern $\Delta'_{2M}$ the automaton 
passes by the fault
and arrives in the state $\widehat{q}$ to some node $v_j$,
with $M+2n+1 < j < M+8n$.
The original computation on the correct pattern $\Delta_{2M}$
regularly visits nodes in this region in the state $\widehat{q}$,
but the node $v_j$ need not be one of those nodes.
The idea is to continue the computation on $\Delta'_{2M}$ from $v_j$,
so that it gets back on the track
of the periodic computation on $\Delta_{2M}$.
For that, the automaton should compensate for the \emph{shift} of $v_j$
relative to the nearest node visited in the state $\widehat{q}$ on $\Delta_{2M}$.
If the shift is zero, then this is all, the Main Lemma is proved.
And if coming with zero shift is impossible,
then there is a proper step in the periodically repeated sequence which can be shrunk,
and the shift will be compensated
by using this shrunken proper step in several iterations of the period.

\begin{proof}[Proof of the Main Lemma.]
First consider the case,
when the automaton can pass by the fault in the faulty pattern $\Delta'_{2M}$
without any shift,
that is, it can move from the configuration $(\widehat{q},v_{i'})$
to a configuration $(\widehat{q},v_j)$, with $M+2n+1 < j < M+8n$,
which is visited in the computation on the correct pattern $\Delta_{2M}$.
Since every iteration of the periodically
repeated sequence is contained in $2n$ consecutive elements $\Delta_2[*,\Delta_a]$,
the automaton will never reach the fault on the faulty pattern again
while making iterations of the period starting from configuration $(\widehat{q},v_j)$.
Then, it can continue the computation on the faulty pattern like on the correct pattern,
and get to configuration $(\widehat{q},v_{i''})$, as desired.
In this case the Main Lemma is proved.

Now, let it be impossible to pass by the fault without a shift.
Then, some proper step of some type $(i,p,q)$
from the periodic part of the computation on the correct pattern 
cannot be reproduced on the faulty pattern.
In this case,
Condition~\ref{claim_shrunk_moves__step_i_on_Delta_prime} in Claim~\ref{claim_shrunk_moves}
does not hold for this step.
Then the claim asserts that
there is a proper step of type $(\widetilde{i},p,q)$,
where $|\widetilde{i}-i|=1$ and $|\widetilde{i}|=|i|-1$.
Let a single iteration of the periodically repeated sequence
move the automaton up by $s$ elements $\Delta_2[*,\Delta_a]$;
and from Claim~\ref{claim_run_Delta_2M_from_v_to_v} it is known that $1 \leqslant s \leqslant n$.
Furthermore, $s \geqslant 2$, because otherwise all nodes from $v_{i'}$ to $v_{i''}$
would be visited in the state $\widehat{q}$,
and the automaton would pass by the fault without a shift,
contrary to the assumption.
Then, if one uses the shrunken proper step of type $(\widetilde{i},p,q)$ in the iteration
instead of the original proper step of type $(i,p,q)$,
then the resulting iteration moves the automaton upwards
either by $s-1$ or by $s+1$ elements $\Delta_2[*,\Delta_a]$
(depending on whether the original proper step was upward or downward).

The idea is to use this iteration with a shrunken step several times to compensate for the shift.
By Claim~\ref{claim_passing_by_v_M_v_M_plus_1},
the automaton can move on the faulty pattern
from $(\widehat{q},v_{i'})$ to $(\widehat{q},v_j)$, with $M+2n+1 < j < M+8n$.
In the computation on the correct pattern the automaton comes to nodes with numbers 
$i'$, $i'+s$, $i'+2s$, \ldots, $i''$ in the state $\widehat{q}$.
And to get back on the track of this computation,
the automaton should compensate for the wrong residue of the number 
$j$ modulo $s$.
This can be done by applying at most $s-1$ 
iterations with the shrunken proper step.
It remains to prove that the automaton cannot reach the fault
and cannot move upward too far
while compensating for the wrong residue modulo $s$.

Since the original iteration has pace at least $2$
and involves at most $2n$ consecutive elements $\Delta_2[*,\Delta_a]$,
the iteration with the shrunken proper step has positive pace
and involves at most $2n+1$ consecutive elements $\Delta_2[*,\Delta_a]$.
Thus, the automaton cannot reach the fault by applying such iterations
starting in the node $v_j$, because $j > M+2n+1$,
and the lowest $2n+1$ nodes possibly visited within these iterations
are from $v_{M+2}$ to $v_{M+2n+2}$.

Now consider how far up the automaton can move while correcting the shift.
It applies at most $s-1$ iterations with the shrunken proper step;
such iterations have pace at most $s+1$.
Therefore, applying these iterations leads the automaton
to a node with number at most $j+(s-1)(s+1)$,
and $j+(s-1)(s+1) < j+s^2 < M+8n+n^2$.
And since each iteration involves at most $2n+1$ consecutive elements $\Delta_2[*,\Delta_a]$,
during these iterations the automaton can visit nodes with numbers 
at most $j+(s-1)(s+1)+2n$, which satisfies $j+(s-1)(s+1)+2n < M+10n+n^2$.
It should be checked that after compensating for the wrong residue modulo $s$ 
the automaton comes to a node with number at most $i''$
(and thus catches up with the periodic part of the computation),
and that it never reaches the root of the faulty pattern.
Therefore, it is enough to check two inequalities:
$M+8n+n^2 \leqslant i''$ and $M+10n+n^2 < 2M$.
The latter inequality holds since $M > n^2+10n$,
and it implies the former inequality because $i'' > 2M-n$.

Therefore, the automaton $A$ can move on the faulty pattern $\Delta_{2M}'$
from configuration $(\widehat{q},v_{i'})$
to configuration $(\widehat{q},v_{i'+ts})$,
for some integer $t$, such that $M+2n+1 < i'+ts \leqslant i''$,
and then, following the computation on the correct pattern,
to configuration $(\widehat{q},v_{i''})$.
Therefore, $(q_{\text{start}},1,q_{\text{finish}},0) \in \delta_{\Delta_{2M}'}$,
and the Main Lemma is proved,
as well as the entire Theorem~\ref{ntwa_rec_L_but_not_compl_of_L}.
\end{proof}

\section{UTWA are weaker than NTWA}\label{section_utwa_weaker_than_ntwa}

An \emph{unambiguous tree-walking automaton (UTWA)}
is a nondeterministic tree-walking automaton
that has at most one accepting computation
on each input tree.
In this section, it is proved that UTWA are strictly weaker than NTWA.

The separating language $L$, defined in Section~\ref{section_separating_language},
is the same as in the rest of this paper.
By Theorem~\ref{ntwa_rec_L_but_not_compl_of_L},
it is recognized by an NTWA $A_L$.

\begin{theorem}\label{theorem_utwa_strictly_weaker_than_ntwa}
The class of tree languages recognized by unambiguous tree-walking automata
is strictly smaller than the class of languages recognized by nondeterministic tree-walking automata.
In particular, no UTWA recognizes the language $L$.
\end{theorem}
\begin{proof}
Let $B$ be any NTWA recognizing the language $L$.
It should be proved that the automaton $B$ is not unambiguous.
Since unambiguity and the language recognized
are not affected by addition of unreachable states,
by Lemma~\ref{symmetries_lemma},
it can be assumed that $B$ is time-symmetric with an even number of states.
Let $n \geqslant 4$ be the number of its states.
Let $\Delta_n$ and $\Delta_{2M}'$, both of rank $1$,
be the patterns defined for the automaton $B$
as in Section~\ref{section_elements_with_and_without_errors}.
By the Main Lemma, $\delta_{\Delta_n} \subseteq \delta_{\Delta_{2M}'}$.

\begin{figure}[tp]
	\centerline{\includegraphics[scale=0.9]{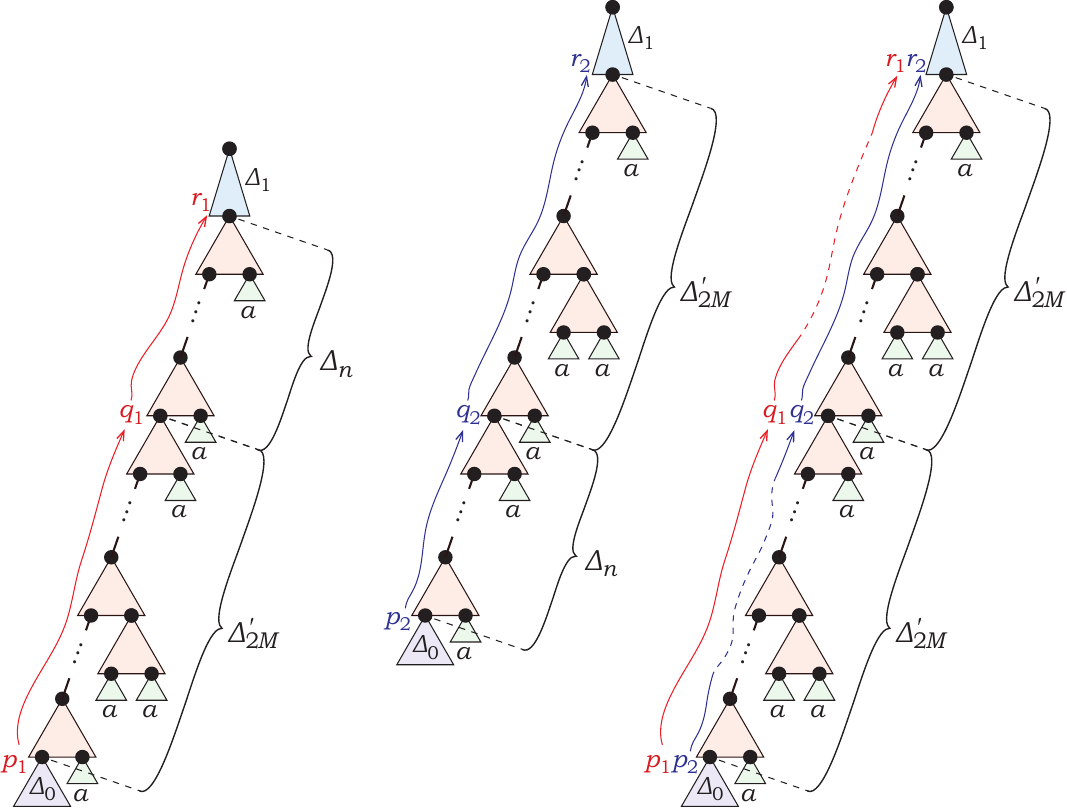}}
	\caption{(left) $T_{0,1}$; (middle) $T_{1,0}$; (right) $T_{1,1}$.}
	\label{f:ntwa_for_L_ambiguous}
\end{figure}

Consider the following four trees.
\begin{align*} 
T_{0,0} &= \mathit{root}[\Delta_1[\Delta_n[\Delta_n[\Delta_0]]]],\\
T_{0,1} &= \mathit{root}[\Delta_1[\Delta_n[\Delta_{2M}'[\Delta_0]]]],\\
T_{1,0} &= \mathit{root}[\Delta_1[\Delta_{2M}'[\Delta_n[\Delta_0]]]],\\
T_{1,1} &= \mathit{root}[\Delta_1[\Delta_{2M}'[\Delta_{2M}'[\Delta_0]]]].
\end{align*}
Of these, the three trees $T_{0,1}, T_{1,0}, T_{1,1}$,
presented in Figure~\ref{f:ntwa_for_L_ambiguous},
are in the language $L$, and hence accepted by the automaton $B$.
The tree $T_{0,0}$ is not in $L$, and must be rejected by $B$.

Let $C_{0,1}$ be an accepting computation on the tree $T_{0,1}$.
By replacing the small correct pattern $\Delta_n$ in the tree $T_{0,1}$
with $\Delta_{2M}'$, and by using the inclusion $\delta_{\Delta_n} \subseteq \delta_{\Delta_{2M}'}$,
the computation $C_{0,1}$ is transformed into an accepting computation $C_{0,1}'$ 
on the tree $T_{1,1}$,
such that for every run through the upper pattern $\Delta_{2M}'$ made in $C'_{0,1}$
there is a run of the same type through the small correct pattern $\Delta_n$ (as it was made in $C_{0,1}$).
This is illustrated in Figure~\ref{f:ntwa_for_L_ambiguous}(left),
where the run from $p_1$ to $r_1$
is transformed to a run on $T_{1,1}$
shown in Figure~\ref{f:ntwa_for_L_ambiguous}(right).

Similarly, an accepting computation $C_{1,0}$ on the tree $T_{1,0}$
is transformed into an accepting computation $C'_{1,0}$ on the tree $T_{1,1}$,
see Figure~\ref{f:ntwa_for_L_ambiguous}(middle,right).
All runs through the lower pattern $\Delta_{2M}'$ made in $C'_{1,0}$
can be reproduced on $\Delta_n$.

Suppose that the accepting computations $C'_{0,1}$ and $C'_{1,0}$ on the tree $T_{1,1}$ are the same.
Then the computation $C'_{0,1}$, while passing through each pattern $\Delta_{2M}'$,
makes only such runs that can be executed on $\Delta_n$.
Replacing both patterns $\Delta_{2M}'$ with $\Delta_n$,
one obtains an accepting computation of the automaton $B$
on the tree $T_{0,0}$, which is not in $L$.
This is a contradiction, and therefore
the computations $C'_{0,1}$ and $C'_{1,0}$ are distinct,
and thus the automaton $B$ is not unambiguous.
\end{proof}

\section{A deterministic one-pebble tree-walking automaton recognizing $L$}

It is worth noting that the language $L$ used in this paper
can be recognized by a deterministic tree-walking automaton with one pebble
(see Boja\'nczyk et al.~\cite{BojanczykSamuelidesSchwentickSegoufin}
for a precise definition of this model).

\begin{theorem}
There is a deterministic tree-walking automaton with one pebble
recognizing the language $L$.
\end{theorem}
\begin{proof}
The automaton moves its pebble in the order of depth-first tree traversal.
When it puts the pebble at some node $v$,
it proceeds with checking the following two conditions:
first, that the left subtree of $v$ contains at least one $a$-labelled leaf,
and secondly, that the right subtree of $v$ contains at least two $a$-labelled leaves.
With the pebble in place,
both searches can be done deterministically.
If both conditions hold for some node $v$,
then the automaton reports that the tree is in $L$.
Otherwise, it moves the pebble to the next position
and begins the next check.

If no node in the tree has both conditions satisfied at the same time,
the automaton eventually completes moving its pebble around
and reports that the tree is not in $L$.
\end{proof}

Since the family of one-pebble deterministic tree-walking automata
is closed under complementation~\cite{MuschollSamuelidesSegoufin},
the complement of the language $L$ is recognized by such an automaton as well.

Thus, the complement of $L$ is a tree language
recognized by a one-pebble deterministic tree-walking automaton,
but not by any NTWA.

\section{Conclusion}

Besides proving the non-closure of nondeterministic tree-walking automata (NTWA)
under complementation,
this paper also provides partially different proofs
of the original results by Boja\'nczyk and Colcombet~\cite{BojanczykColcombet_det,BojanczykColcombet_reg}.
First, since deterministic tree-walking automata (DTWA) are closed under complementation,
the result of this paper implies that NTWA cannot be determinized~\cite{BojanczykColcombet_det}.
Secondly, bottom-up tree-walking automata are closed under complementation as well,
and hence they are stronger than NTWA~\cite{BojanczykColcombet_reg}.

The second result of this paper, that is,
that unambiguous tree-walking automata (UTWA) are weaker than nondeterministic ones,
leaves a few related questions to investigate.
Most importantly, it remains unknown whether UTWA
are any more powerful than DTWA.
If these families turn out to be different,
then another question will arise,
whether UTWA are closed under complementation.
Furthermore, the same questions can be asked about \emph{unambiguous graph-walking automata} (UGWA):
if they can be determinized, then so can be UTWA,
but it could also be possible that UTWA can be determined whereas UGWA cannot.

\section*{Acknowledgement}

This work was supported by the Russian Science Foundation, project 23-11-00133.

\end{document}